\documentclass[aps,twocolumn]{revtex4-1}
\usepackage[utf8]{inputenc} 

\usepackage{times}
\usepackage{graphicx}
\usepackage{amsfonts}
\usepackage{amsmath, amsthm, amssymb}
\usepackage{dsfont}
\usepackage{color}
\usepackage{empheq}
\usepackage{standalone}
\usepackage[most]{tcolorbox}
\usepackage{physics}
\usepackage{bm}
\usepackage{amssymb,amsmath,amsfonts,amscd}
\usepackage{mathrsfs}
\usepackage{siunitx}
\usepackage{commath}
\usepackage{graphicx}       
\graphicspath{ {Images/} }  
\usepackage{braket}
\usepackage{bm}
\usepackage{empheq}
\usepackage[most]{tcolorbox}
\usepackage{physics}
\newcommand*{\vrec}{\,{\setlength{\fboxsep}{1pt}\fbox{\phantom{l}}}\,}

\usepackage{hyperref} 
\hypersetup{
    colorlinks=true,
    linkcolor=blue,    
    citecolor=blue,    
    urlcolor=blue,      
}

\begin{document}

\title{Schwinger boson study of superconductivity mediated by antiferromagnetic spin-fluctuations}

\author{Eirik Erlandsen}
\affiliation{\mbox{Center for Quantum Spintronics, Department of Physics, Norwegian University of Science and Technology,}\\NO-7491 Trondheim, Norway}
 
\author{Asle Sudbø}
\email[Corresponding author: ]{asle.sudbo@ntnu.no}
\affiliation{\mbox{Center for Quantum Spintronics, Department of Physics, Norwegian University of Science and Technology,}\\NO-7491 Trondheim, Norway}

\begin{abstract}
We study superconductivity in a normal metal, arising from effective electron-electron interactions mediated by spin-fluctuations in a neighboring antiferromagnetic insulator. Introducing a frustrating next-nearest neighbor interaction in a Néel antiferromagnet with an uncompensated interface, the superconducting critical temperature is found to be enhanced as the frustration is increased. Further, for sufficiently large next-nearest neighbor interaction, the antiferromagnet is driven into a stripe phase, which can also give rise to attractive electron-electron interactions. For the stripe phase, as previously reported for the Néel phase, the superconducting critical temperature is found to be amplified for an uncompensated interface where the normal metal conduction electrons are coupled to only one of the two sublattices of the magnet. The superconducting critical temperature arising from fluctuations in the stripe phase antiferromagnet can be further enhanced by approaching the transition back to the Néel phase.   
\end{abstract}

\pacs{Valid PACS appear here}

\maketitle

\section{Introduction}
Recent studies have investigated whether spin fluctuations in a magnetic material can induce attractive interactions between electrons in an adjacent conductor, leading to a superconducting instability \cite{Kargarian2016, Gong2017, Fjaerbu2018, Hugdal2018, Fjaerbu2019, Erlandsen2019, Erlandsen2020}. Both ferromagnetic (FMIs) and antiferromagnetic insulators (AFMIs) have been considered as potential sources for the magnetic fluctuations. In order to ensure magnetic ordering, these materials typically ought to be of three-dimensional nature \cite{Mermin1966, Hohenberg1967}. Since a three-dimensional crystal has more than one crystal plane, the issue arises whether it makes a difference which crystal plane is exposed at the interface. For the simplest type of ferromagnet, all lattice sites can be considered to be identical, and the choice of crystal plane does not affect the coupling to an external system. For the simplest case of an antiferromagnet on a bipartite lattice, there are two different types of lattice sites. Depending on the chosen crystal plane, it is possible that either both sublattices (compensated interface) are exposed at the interface, or only one of the sublattices (uncompensated interface) is exposed at the interface \cite{Nogues1999, Nogues2005, Stamps2000}.\\
\indent As outlined in Ref.\! \cite{Kamra2019}, the presence of an antiferromagnetic eigen-excitation with spin unity is associated with a large, and oppositely directed, spin located on each of the two sublattices. An external system that is only coupled to one of the two sublattices is then essentially interacting with a large spin, potentially leading to a strong coupling interaction. In accordance with this picture, uncompensated antiferromagnetic interfaces have been predicted to enhance the spin transfer to a neighboring conductor \cite{Kamra2017}, and produce magnon-mediated indirect exciton condensation \cite{Johansen2019}. Moreover, importantly for our purposes, coupling a conductor to an uncompensated, instead of compensated, antiferromagnetic interface might produce a stronger induced electron-electron interaction and higher superconducting critical temperature \cite{Erlandsen2019, Erlandsen2020}.\\ 
\indent In view of the fact that the superconductivity arises from magnetic fluctuations, it is natural to ask whether amplifying the fluctuations can be favorable. One way of achieving such an amplification is to include next-nearest neighbor frustration in the AFMI. This type of frustration is common in antiferromagnets, and has been predicted to increase the critical temperature of superconductivity induced on the surface of a topological insulator \cite{Erlandsen2020}. Using the picture from Ref.\! \cite{Kamra2019}, this increase in critical temperature can be understood from the amplified fluctuations increasing the average spin on each sublattice associated with an antiferromagnetic magnon. The effect of coupling to only one of the two sublattices then becomes stronger.\\
\indent The previous study of the effect of frustration did, however, employ a Holstein-Primakoff treatment of the AFMI, starting from a staggered Néel state \cite{Erlandsen2020}. The study was therefore limited to the case of small antiferromagnetic next-nearest neighbor exchange coupling $J_2$ compared to the nearest neighbor coupling $J_1$. Hence, it is of interest to investigate the rest of the phase diagram of the $J_1$-$J_2$ Heisenberg model. On a square or cubic lattice, this model contains two distinct magnetically ordered phases, a Néel phase for $J_2/J_1 \ll 1$ and a stripe phase for $J_2/J_1 \gg 1$ \cite{Parkinson2010, Farnell2016}. In the stripe phase, the spins in e.g.\! one column could be aligned with each other and anti-aligned with the spins in the neighboring columns, creating alternating stripes of spins. This state arises from two decoupled, interpenetrating, Néel ordered antiferromagnets ($J_1 = 0$), which align themselves and create a stripe pattern for finite $J_1$ \cite{Flint2009, Henley1989}. Given the origin of the stripe phase, it could be possible that coupling to only the up/down spins of both Néel ordered antiferromagnets could give a similar effect as only coupling to one of the sublattices of a single Néel ordered antiferromagnet.\\
\indent The transition between the Néel and stripe phases takes place in the vicinity of $J_2/J_1 = 0.5$, with variations depending on the spin quantum number and lattice structure \cite{Parkinson2010, Farnell2016, Haghshenas2018, Bishop_2008}. For the spin-$1/2$ system on a square lattice, it is predicted that there is an intermediate region where the magnetic long-range order is destroyed by quantum fluctuations \cite{Parkinson2010, Dagotto1989, Figueirido1990, Capriotti2001}. A similar intermediate region might also be present for the spin-$1/2$ system on a simple cubic lattice, in contrast to the case of a body-centered cubic lattice \cite{Schmidt2002, Farnell2016}. We will, however, focus on the properties of the ordered phases. While the Néel phase is more commonly encountered, the stripe configuration has attracted attention as the magnetic ground state of the iron oxypnictide $\rm{LaOFeAs}$, which is the original undoped parent compound of the high-$T_c$ iron pnictides \cite{Kamihara2008}. This layered material has been found to be well described by the square lattice $J_1$-$J_2$ Heisenberg model with spin $S>1/2$ and $J_2/J_1 > 1/2$ \cite{Si2008, Ma2008}.\\
\indent In this paper, we consider an AFMI, with both nearest-neighbor and next-nearest neighbor antiferromagnetic exchange interaction, which is proximity-coupled to a normal metal (NM). The AFMI interface can be either compensated or uncompensated. In order to better take into account the effect of the frustration, we perform a Schwinger boson study, rather than the usual Holstein-Primakoff treatment which has been employed for these systems in the past. Further, we also extend the analysis into the regime where the frustration drives the AFMI into a stripe phase. The two subsystems are coupled through an interfacial exchange coupling, which produces effective electron-electron interactions in the NM. We explore the effect of the induced interactions through a BCS-type mean-field treatment, and numerically solve the gap equation in order to determine how the critical temperature depends on the properties of the AFMI.\\ 
\indent For a Néel AFMI with small next-nearest neighbor frustration, the results for the superconductivity are similar to the results obtained through a Holstein-Primakoff treatment of the AFMI \cite{Erlandsen2019}. As before, the strength of the effective interactions is enhanced for an uncompensated interface, leading to an amplified critical temperature. Increasing the frustration, the effect of coupling to only one of the two sublattices of the AFMI becomes stronger, as expected \cite{Erlandsen2020}. Further, the increased frustration also lowers the cutoff on the boson spectrum, and reduces the sublattice magnetization in the AFMI, which is found to reduce the strength of the induced electron-electron interactions. The overall effect is however still, typically, a rise in the critical temperature when the frustration is increased. For the stripe phase, coupling to an uncompensated AFMI interface is found to enhance the critical temperature, like in the Néel case. Moreover, approaching the transition point between the two magnetic phases from the stripe side does, like from the Néel side, lead to a further increase in the critical temperature.\\
\indent The paper is organized as follows. In Sec.\! \ref{Section:Model} we introduce the modelling of the system. In Sec.\! \ref{Section:AFMI} the Schwinger boson treatment of the antiferromagnet is covered for both the Néel phase and the stripe phase. Next, the NM and the coupling between the two subsystems is treated in Sec.\! \ref{Section:Coupling}. In Sec.\! \ref{Section:Eff_int}, we derive an effective theory of interacting electrons, and in Sec.\! \ref{Section:Gap_equation} we investigate the possibility of a superconducting instability through a weak-coupling mean-field theory. The results from the numerical treatment of the gap equation is presented in Sec.\! \ref{Section:Results}. Finally, in Sec.\! \ref{Section:Summary}, we summarize our results. Additional details concerning the derivation of the interaction potential is included in the Appendix. 

\section{Model}\label{Section:Model}
\begin{figure}[t] 
    \begin{center}
        \includegraphics[width=0.95\columnwidth,trim= 2.5cm 18.2cm 4.0cm 1.8cm,clip=true]{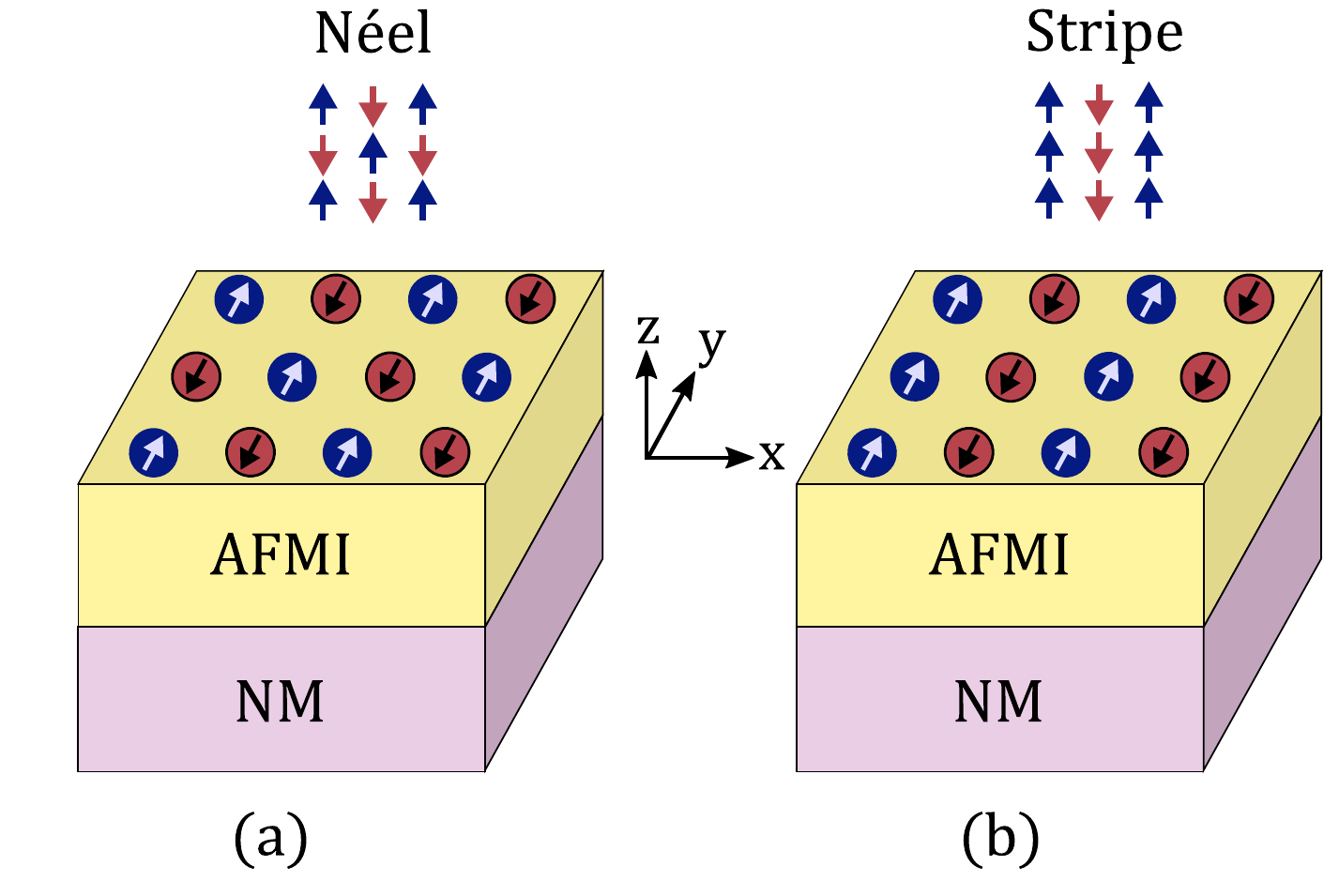}
    \end{center}
    \caption{The system consists of a normal metal (NM), which is proximity coupled to an antiferromagnetic insulator (AFMI). The AFMI can be either in a Néel phase (a) or a stripe phase (b).}
    \label{fig:System}
\end{figure} 
\indent
The system, consisting of a NM proximity-coupled to an AFMI, is displayed in Fig.\! \ref{fig:System}. The real system we have in mind would consist of a three dimensional AFMI grown on top of a thin NM layer. However, in order to capture the physics at the interface, we apply two-dimensional lattice models, with continuous boundary conditions, for the two subsystems. The AFMI is described by a Heisenberg Hamiltonian with nearest neighbor and next-nearest neighbor exchange interaction, as well as easy-axis anisotropy ($K$). By tuning the next-nearest neighbor interaction, the ground state of the AFMI can then be changed from a Néel phase to a stripe phase. The NM is described by a tight-binding hopping model. The two subsystems are coupled together through an interfacial exchange coupling ($\bar{J}$) where the spins of the NM conduction electrons are coupled to the AFMI lattice site spins \cite{Kajiwara2010, Takahashi2010, Bender2015, Fjaerbu2018, Fjaerbu2019}.\\
\indent For each of the magnetic phases we define two sublattices. In Fig.\! \ref{fig:System}, the lattice sites with blue spins constitute one sublattice, and the lattice sites with oppositely aligned red spins constitute the other sublattice. The sublattices are therefore defined differently for the Néel phase and the stripe phase. As mentioned earlier, depending on which crystal plane of the three dimensional AFMI that constitutes the interface, it is possible that either both sublattices or only one of the sublattices is exposed at the interface. In order to describe this, we apply a model where the NM electrons are coupled to both sublattices of the AFMI, but where the coupling strength is allowed to differ for the two sublattices ($\bar{J}_A/\bar{J}_B$) \cite{Erlandsen2019}. This model allows us to tune between the two cases of a compensated or uncompensated interface.\\
\indent The system is modelled by a Hamiltonian $H = H_{\rm{AFMI}} + H_{\text{NM}} + H_{\text{int}}$ where
\begin{subequations}
\begin{align}
    &H_{\rm{AFMI}} = J_1 \!\sum_{\langle \bm{i}, \bm{j} \rangle}\bm{S}_{\bm{i}}\cdot \bm{S}_{\bm{j}} + J_2\! \sum_{\langle\langle \bm{i}, \bm{j} \rangle\rangle}\!\bm{S}_{\bm{i}}\cdot \bm{S}_{\bm{j}} - K\sum_{\bm{i}}S^2_{\bm{i}z},\\
    &H_{\text{NM}} =  -t\!\sum_{\langle \bm{i}, \bm{j} \rangle\sigma} c^{\dagger}_{\bm{i}\sigma}c_{\bm{j}\sigma} - \mu \sum_{\bm{i}\sigma}c^{\dagger}_{\bm{i}\sigma}c_{\bm{i}\sigma},\\
    &H_{\text{int}} = -2\Bar{J}_{A}\sum_{\bm{i}\in A}c_{\bm{i}}^{\dagger}\bm{\tau}c_{\bm{i}}\cdot \bm{S_i} -2\Bar{J}_{B}\sum_{\bm{i}\in B}c_{\bm{i}}^{\dagger}\bm{\tau}c_{\bm{i}}\cdot \bm{S_i}. 
\end{align}
\end{subequations}
Here, $c_{\bm{i}}^{\dagger} = (c_{\bm{i}\uparrow}^{\dagger}, c_{\bm{i}\downarrow}^{\dagger})$ and $c_{\bm{i}\sigma}^{\dagger}$ creates an electron with spin $\sigma$ on lattice site $\bm{i}$. The electron hopping amplitude is denoted by $t$, and $\mu$ is the chemical potential. The easy-axis anisotropy constant $K$ is taken to be positive, favoring ordering of spins in the $z$-direction in spin space, which could be either parallel with or normal to the interface. In the interaction part of the Hamiltonian, $\bm{\tau}$ is the vector of Pauli matrices, representing the NM electron spin which is coupled to the lattice site spin $\bm{S}_{\bm{i}}$ in the AFMI. Further, it should be noted that the lattices are assumed to be square, the sums over nearest and next-nearest neighbors include the neighbors in both positive and negative spatial directions, and we have set $\hbar = a = 1$, where $a$ is the lattice constant.  

\section{Antiferromagnet}\label{Section:AFMI}

In order to treat the AFMI, we will represent the lattice site spins in terms of Schwinger bosons \cite{Arovas1988, auerbach_book, Sarker1989}. For our purposes, where we will couple an external system to the two sublattices of the AFMI, it will be useful to define different Schwinger bosons for the two sublattices $A$ and $B$ \cite{Wei1994}

\begin{subequations}
\begin{align}
    &S^{A}_{\bm{i}+} = a^{\dagger}_{\bm{i}\uparrow} a_{\bm{i}\downarrow},\\
    &S^{A}_{\bm{i}-} = a^{\dagger}_{\bm{i}\downarrow} a_{\bm{i}\uparrow},\\
    &S^{A}_{\bm{i}z} = \frac{1}{2}(a^{\dagger}_{\bm{i}\uparrow}a_{\bm{i}\uparrow} - a^{\dagger}_{\bm{i}\downarrow}a_{\bm{i}\downarrow}),
\end{align}
\end{subequations}

\begin{subequations}
\begin{align}
    &S^{B}_{\bm{i}+} = -b^{\dagger}_{\bm{i}\downarrow} b_{\bm{i}\uparrow},\\
    &S^{B}_{\bm{i}-} = -b^{\dagger}_{\bm{i}\uparrow} b_{\bm{i}\downarrow},\\
    &S^{B}_{\bm{i}z} = -\frac{1}{2}(b^{\dagger}_{\bm{i}\uparrow}b_{\bm{i}\uparrow} - b^{\dagger}_{\bm{i}\downarrow}b_{\bm{i}\downarrow}).
\end{align}
\end{subequations}
An ordered Néel or stripe state can then be described through a condensation \cite{auerbach_book, Sarker1989, Powell2014} of $\uparrow$-bosons with momentum $\bm{k} = 0$ on both the $A$ and $B$ sublattice, producing a spatially uniform state with opposite magnetization on the two sublattices. In order to fix the length of the spins, we have the condition \cite{auerbach_book}

\begin{align}
\begin{aligned}
    &n_{\bm{i},A} = \sum_{\alpha}a^{\dagger}_{\bm{i}\alpha}a_{\bm{i}\alpha} = 2S,\\
    &n_{\bm{j},B} = \sum_{\alpha}b^{\dagger}_{\bm{j}\alpha}b_{\bm{j}\alpha} = 2S,
    \label{eq:constraint}
\end{aligned}
\end{align}
on each lattice site. In the following mean-field treatment, this condition on the number of Schwinger bosons will be enforced on the average. In order to rewrite the AFMI Hamiltonian in terms of Schwinger boson operators, we introduce bond operators quadratic in the boson operators. We follow the recipe of Ref.\! \cite{Ceccatto1993}, as outlined in Ref.\! \cite{Powell2014} and \cite{Bauer2017}. When the Schwinger boson operators have been defined equally on all lattice sites, the bond operators then take the form

\begin{subequations}
\begin{align}
    A_{\bm{i}\bm{j}} &= \frac{1}{2}\big(a_{\bm{i}\uparrow}a_{\bm{j}\downarrow} - a_{\bm{i}\downarrow}a_{\bm{j}\uparrow}\big),\\
    B_{\bm{i}\bm{j}} &= \frac{1}{2}\big(a_{\bm{i}\uparrow}a^{\dagger}_{\bm{j}\uparrow} + a_{\bm{i}\downarrow}a^{\dagger}_{\bm{j}\downarrow}\big).
\end{align}
\end{subequations}
Here, $A_{\bm{i}\bm{j}}$ corresponds to an antiferromagnetic bond and $B_{\bm{i}\bm{j}}$ corresponds to a ferromagnetic bond \cite{Flint2009, Powell2014}. This choice of bond operators captures the cost of frustrating bonds, which is essential for frustrated antiferromagnets, and has been shown to preserve the time-inversion properties of the spins \cite{Flint2009}. As we have defined different Schwinger boson operators on the two sublattices, we should perform the following transformation on the operators living on the $B$-sublattice in the above definitions of the bond operators 
\begin{align*}
    &a_{\bm{i}\uparrow} \rightarrow - b_{\bm{i}\downarrow},\\
    &a_{\bm{i}\downarrow} \rightarrow b_{\bm{i}\uparrow}.
\end{align*}

\subsection{Néel phase}

For the Néel phase, the AFMI Hamiltonian can be expressed on the form 

\begin{align}
\begin{aligned}
    H^{\text{Néel}}_{\rm{AFMI}} &= J_1 \sum_{\substack{\bm{i} \in A\\\bm{j} \text{nn} \bm{i}}}\bm{S}^{A}_{\bm{i}}\cdot \bm{S}^{B}_{\bm{j}} + J_1 \sum_{\substack{\bm{i} \in B\\\bm{j} \text{nn} \bm{i}}}\bm{S}^{B}_{\bm{i}}\cdot \bm{S}^{A}_{\bm{j}}\\
    &+ J_2\! \sum_{\substack{\bm{i} \in A\\\bm{j} \text{nnn} \bm{i}}}\bm{S}^{A}_{\bm{i}}\cdot \bm{S}^{A}_{\bm{j}} + J_2\! \sum_{\substack{\bm{i} \in B\\\bm{j} \text{nnn} \bm{i}}}\bm{S}^{B}_{\bm{i}}\cdot \bm{S}^{B}_{\bm{j}}\\
    &- K\sum_{\bm{i}\in A}S^2_{\bm{i}z} - K\sum_{\bm{i}\in B}S^2_{\bm{i}z}.
\end{aligned}
\end{align}

We then introduce the bond operators 

\begin{subequations}
\begin{align}
    A^{1,A}_{\bm{i}\bm{j}} &= \frac{1}{2}\big(a_{\bm{i}\uparrow}b_{\bm{j}\uparrow} + a_{\bm{i}\downarrow}b_{\bm{j}\downarrow}\big),\\
    B^{1,A}_{\bm{i}\bm{j}} &= \frac{1}{2}\big(a_{\bm{i}\downarrow}b^{\dagger}_{\bm{j}\uparrow} - a_{\bm{i}\uparrow}b^{\dagger}_{\bm{j}\downarrow}\big),\\
    A^{1,B}_{\bm{i}\bm{j}} &= \frac{1}{2}\big(-b_{\bm{i}\downarrow}a_{\bm{j}\downarrow} - b_{\bm{i}\uparrow}a_{\bm{j}\uparrow}\big),\\
    B^{1,B}_{\bm{i}\bm{j}} &= \frac{1}{2}\big(b_{\bm{i}\uparrow}a^{\dagger}_{\bm{j}\downarrow} - b_{\bm{i}\downarrow}a^{\dagger}_{\bm{j}\uparrow}\big),\\
    A^{2,A}_{\bm{i}\bm{j}} &= \frac{1}{2}\big(a_{\bm{i}\uparrow}a_{\bm{j}\downarrow} - a_{\bm{i}\downarrow}a_{\bm{j}\uparrow}\big),\\
    B^{2,A}_{\bm{i}\bm{j}} &= \frac{1}{2}\big(a_{\bm{i}\uparrow}a^{\dagger}_{\bm{j}\uparrow} + a_{\bm{i}\downarrow}a^{\dagger}_{\bm{j}\downarrow}\big),\\
    A^{2,B}_{\bm{i}\bm{j}} &= \frac{1}{2}\big(b_{\bm{i}\uparrow}b_{\bm{j}\downarrow} - b_{\bm{i}\downarrow}b_{\bm{j}\uparrow}\big),\\
    B^{2,B}_{\bm{i}\bm{j}} &= \frac{1}{2}\big(b_{\bm{i}\uparrow}b^{\dagger}_{\bm{j}\uparrow} + b_{\bm{i}\downarrow}b^{\dagger}_{\bm{j}\downarrow}\big),
\end{align}
\end{subequations}
and write out the Hamiltonian as

\begin{align}
\begin{aligned}
    H^{\text{Néel}}_{\rm{AFMI}} &= J_1 \sum_{\substack{\bm{i} \in A\\\bm{j} \text{nn} \bm{i}}}\Big[\big(B^{1,A}_{\bm{i}\bm{j}}\big)^{\dagger}B^{1,A}_{\bm{i}\bm{j}} - \big(A^{1,A}_{\bm{i}\bm{j}}\big)^{\dagger}A^{1,A}_{\bm{i}\bm{j}} - \frac{1}{4}n_{\bm{i},A}\Big]\\
    &+ J_1 \sum_{\substack{\bm{i} \in B\\\bm{j} \text{nn} \bm{i}}}\Big[\big(B^{1,B}_{\bm{i}\bm{j}}\big)^{\dagger}B^{1,B}_{\bm{i}\bm{j}} - \big(A^{1,B}_{\bm{i}\bm{j}}\big)^{\dagger}A^{1,B}_{\bm{i}\bm{j}} - \frac{1}{4}n_{\bm{i},B}\Big]\\
    &+ J_2\! \sum_{\substack{\bm{i} \in A\\\bm{j} \text{nnn} \bm{i}}}\Big[\big(B^{2,A}_{\bm{i}\bm{j}}\big)^{\dagger}B^{2,A}_{\bm{i}\bm{j}} - \big(A^{2,A}_{\bm{i}\bm{j}}\big)^{\dagger}A^{2,A}_{\bm{i}\bm{j}} - \frac{1}{4}n_{\bm{i},A}\Big]\\
    &+ J_2\! \sum_{\substack{\bm{i} \in B\\\bm{j} \text{nnn} \bm{i}}}\Big[\big(B^{2,B}_{\bm{i}\bm{j}}\big)^{\dagger}B^{2,B}_{\bm{i}\bm{j}} - \big(A^{2,B}_{\bm{i}\bm{j}}\big)^{\dagger}A^{2,B}_{\bm{i}\bm{j}} - \frac{1}{4}n_{\bm{i},B}\Big]\\
    &- K\sum_{\bm{i}\in A}S^2_{\bm{i}z} - K\sum_{\bm{i}\in B}S^2_{\bm{i}z} + \lambda_A \sum_{\bm{i} \in A}\big(n_{\bm{i},A}- \kappa\big)\\
    &+ \lambda_B\sum_{\bm{i} \in B}\big(n_{\bm{i},B} - \kappa\big).
\end{aligned}
\end{align}
Here $\kappa = 2S$ and $\lambda_A$, $\lambda_B$ are Lagrange multipliers that have been included in order to enforce the constraint on the number of Schwinger bosons per site. The choice of $\kappa = 2S$ seems sensible based on equation \eqref{eq:constraint}, and fixes the magnitude of the spins to the correct value. For this value of $\kappa$, the spin-fluctuations are, however, somewhat over-estimated \cite{auerbach_book}. Another possibility is therefore to adjust $\kappa$ in order to obtain the correct result for the fluctuations, at the expense of the spin length \cite{Messio2013}. We will mostly be interested in how the results vary depending on $J_2$ for typical values of the rest of the parameters, and the specific choice of $\kappa$ therefore is not of great importance.\\
\indent We next perform a mean-field decoupling of a  bond-variable $C_{ij}$ as follows
\begin{align}
\begin{aligned}
    &C_{\bm{i}\bm{j}} =  \langle C_{\bm{i}\bm{j}} \rangle + \Big(C_{\bm{i}\bm{j}} -  \langle C_{\bm{i}\bm{j}} \rangle \Big) \equiv \langle C_{\bm{i}\bm{j}} \rangle + \delta(C_{\bm{i}\bm{j}}),\\
    &\big(C_{\bm{i}\bm{j}}\big)^{\dagger} C_{\bm{i}\bm{j}} \approx \langle C_{\bm{i}\bm{j}} \rangle^{\dagger}C_{\bm{i}\bm{j}} + \langle C_{\bm{i}\bm{j}} \rangle\big(C_{\bm{i}\bm{j}}\big)^{\dagger} -\abs{\langle C_{\bm{i}\bm{j}} \rangle}^2.
    \label{eq:C}
\end{aligned}
\end{align}
Here, we have neglected quadratic terms in the deviations from the mean-field values. Moreover, we choose an Ansatz for the expectation values of the bond operators that will produce a Néel type state 

\begin{subequations}
\begin{align*}
    &\langle B^{1,A}_{\bm{i}\bm{j}} \rangle  = \langle B^{1,B}_{\bm{i}\bm{j}} \rangle = 0,\\
    &\langle A^{1,A}_{\bm{i}\bm{j}} \rangle  = -\langle A^{1,B}_{\bm{i}\bm{j}} \rangle \equiv \mathcal{A}_{\bm{\delta}_1},\\
    &\langle B^{2,A}_{\bm{i}\bm{j}} \rangle  = \langle B^{2,B}_{\bm{i}\bm{j}} \rangle \equiv \mathcal{B}_{\bm{\delta}_2},\\
    &\langle A^{2,A}_{\bm{i}\bm{j}} \rangle  = \langle A^{2,B}_{\bm{i}\bm{j}} \rangle = 0,
\end{align*}
\end{subequations}
where all quantities are assumed to be real \cite{Bauer2017, Messio2013}. We also take $\lambda_A = \lambda_B \equiv \lambda$. For the easy-axis terms we do the same mean-field treatment as above and take $\langle S^{C}_{\bm{i}z}\rangle \equiv m_{C}$.\\
\indent We introduce Fourier transformations for the Schwinger boson operators
\begin{figure}[t] 
    \begin{center}
        \includegraphics[width=0.95\columnwidth,trim= 3.2cm 21.5cm 5.0cm 0.7cm,clip=true]{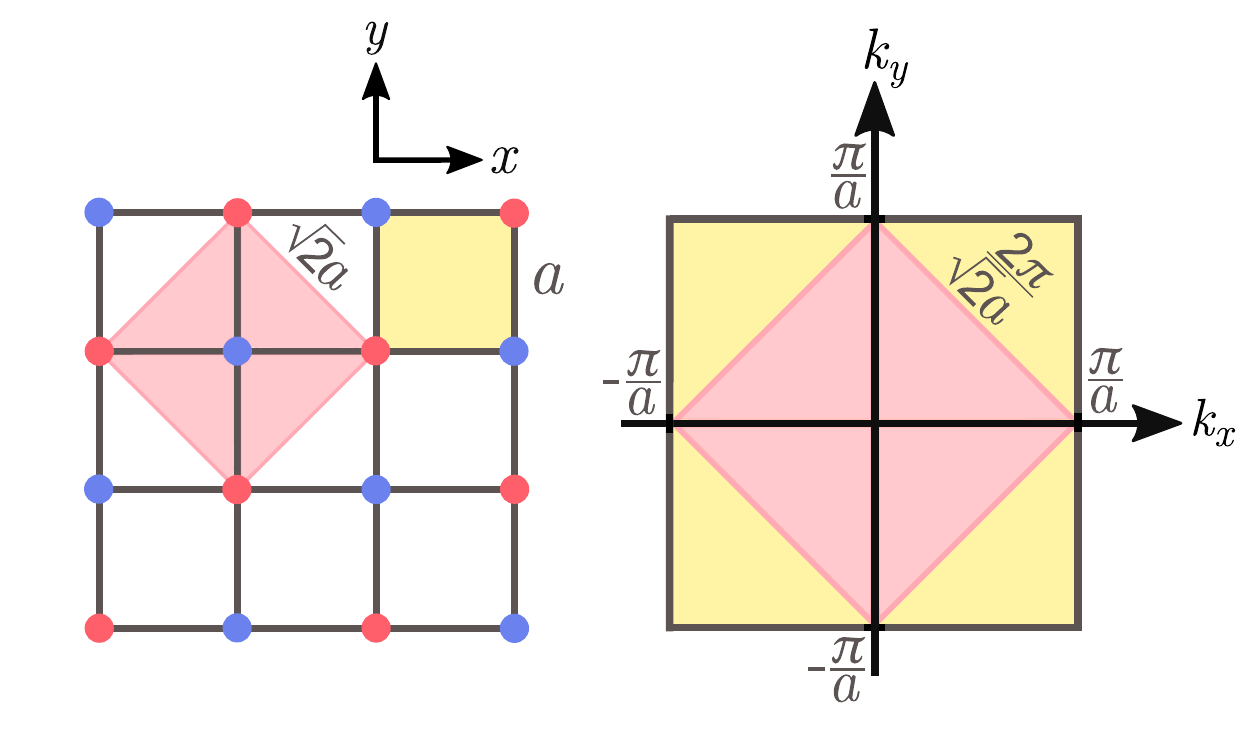}
    \end{center}
    \caption{Unit cell and Brillouin zone for the full lattice and the sublattices.}
    \label{fig:Neel_k}
\end{figure} 
\begin{subequations}
\begin{align}
    &a_{\bm{i}\sigma} = \frac{1}{\sqrt{N_A}}\sum_{\bm{k}\in\diamondsuit}e^{i\bm{k}\cdot \bm{r}_{\bm{i}}}a_{\bm{k}\sigma},\\  &b_{\bm{i}\sigma} = \frac{1}{\sqrt{N_B}}\sum_{\bm{k}\in\diamondsuit}e^{i\bm{k}\cdot \bm{r}_{\bm{i}}}b_{\bm{k}\sigma},
\end{align}
\end{subequations}
where the momenta live in the reduced Brillouin zone of the sublattices $\diamondsuit$, as displayed in Fig.\! \ref{fig:Neel_k}. The AFMI Hamiltonian then takes the following form
\begin{align}
\begin{aligned}
    &H^{\text{Néel}}_{\rm{AFMI}} = 2 N_A \Big[J_1 \sum_{\bm{\delta}_1}\big(\mathcal{A}_{\bm{\delta}_1}\big)^2 - J_2 \sum_{\bm{\delta}_2}\big(\mathcal{B}_{\bm{\delta}_2}\big)^2\Big]\\
    &+ \Big[\lambda - \frac{1}{4}\big(J_1 z_1 + J_2 z_2\big)\Big ]\sum_{\substack{\bm{k}\in\diamondsuit\\\sigma}}\big(a^{\dagger}_{\bm{k}\sigma}a_{\bm{k}\sigma} + b^{\dagger}_{\bm{k}\sigma}b_{\bm{k}\sigma}\big)\\
    & - K m_A \sum_{\substack{\bm{k}\in\diamondsuit\\ \sigma}}\sigma a^{\dagger}_{\bm{k}\sigma}a_{\bm{k}\sigma} + K m_B\sum_{\substack{\bm{k}\in\diamondsuit\\ \sigma}}\sigma b^{\dagger}_{\bm{k}\sigma}b_{\bm{k}\sigma}\\
    & + \sum_{\substack{\bm{k}\in\diamondsuit\\ \sigma}}\gamma^{B_2}_{\bm{k}}\big( a^{\dagger}_{\bm{k}\sigma}a_{\bm{k}\sigma} + b_{\bm{k}\sigma}b^{\dagger}_{\bm{k}\sigma}\big) + K N_A( m^2_A + m^2_B)\\
    &- \sum_{\substack{\bm{k}\in\diamondsuit\\ \sigma}}\gamma^{A_1}_{\bm{k}}\big(b^{\dagger}_{\bm{k}\sigma}a^{\dagger}_{-\bm{k}\sigma} + a_{\bm{k}\sigma}b_{-\bm{k}\sigma}\big) - 2 N_A \kappa \lambda.
\end{aligned}
\end{align}
Here, we have introduced the form-factors 

\begin{subequations}
\begin{align}
    &\gamma^{A_1}_{\bm{k}} \equiv J_1\! \sum_{\bm{\delta}_1}\mathcal{A}_{\bm{
    \delta}_1}  \cos(\bm{k}\cdot \bm{\delta_1}),\\
    &\gamma^{B_2}_{\bm{k}} \equiv J_2\! \sum_{\bm{\delta}_2}\mathcal{B}_{\bm{
    \delta}_2}  \cos(\bm{k}\cdot \bm{\delta_2}),
\end{align}
\end{subequations}
where the sums over nearest neighbors ($\bm{\delta}_1$) and next-nearest neighbors ($\bm{\delta}_2$) cover both positive and negative directions. We then define $\lambda' \equiv \lambda - \frac{1}{4}\big(J_1 z_1 + J_2 z_2\big)$, rename $\lambda' \rightarrow \lambda$, and neglect constant terms that do not contain any of the mean-field parameters.\\
\indent In order to make progress, we split the Hamiltonian up into three parts $H^{\text{Néel}}_{\rm{AFMI}} = E_0 + H^{\uparrow} + H^{\downarrow}$ where

\begin{align}
\begin{aligned}
    E_0 &= 2 N_A \Big[J_1 \sum_{\bm{\delta}_1}\big(\mathcal{A}_{\bm{\delta}_1}\big)^2 - J_2 \sum_{\bm{\delta}_2}\big(\mathcal{B}_{\bm{\delta}_2}\big)^2\Big]\\
    &- 2 N_A\lambda(\kappa  + 1)  + K N_A( m^2_A + m^2_B),
\end{aligned}
\end{align}
and

\begin{align}
\begin{aligned}
    H^{\sigma} &= \sum_{\bm{k}\in\diamondsuit}\big(\lambda + \gamma^{B_2}_{\bm{k}} - \sigma K m_A \big)a^{\dagger}_{\bm{k}\sigma}a_{\bm{k}\sigma}\\
    &+ \sum_{\bm{k}\in\diamondsuit}\big(\lambda + \gamma^{B_2}_{\bm{k}} + \sigma K m_B\big)b_{\bm{k}\sigma}b^{\dagger}_{\bm{k}\sigma}\\
    &- \sum_{\bm{k}\in\diamondsuit}\gamma^{A_1}_{\bm{k}}\big(b^{\dagger}_{\bm{k}\sigma}a^{\dagger}_{-\bm{k}\sigma} + a_{\bm{k}\sigma}b_{-\bm{k}\sigma}\big).
\end{aligned}
\end{align}
We can then perform a Bogoliubov transformation

\begin{align}
\begin{aligned}
    a_{\bm{k}\sigma} &= u_{\bm{k}\sigma} \alpha_{\bm{k}\sigma} - v_{\bm{k}\sigma} \beta^{\dagger}_{-\bm{k}\sigma},\\
    b^{\dagger}_{-\bm{k}\sigma} &= v_{\bm{k}\sigma} \alpha_{\bm{k}\sigma} -  u_{\bm{k}\sigma}\beta^{\dagger}_{-\bm{k}\sigma}. 
    \label{eq:bogoliubov}
\end{aligned}
\end{align}
where $u_{\bm{k}\sigma}$ and $v_{\bm{k}\sigma}$ are taken to be real and are parametrized by $u_{\bm{k}\sigma} = \cosh(\theta_{\bm{k}\sigma})$, $v_{\bm{k}\sigma} = \sinh(\theta_{\bm{k}\sigma})$. The value of $\theta_{\bm{k}\sigma}$ that diagonalizes the Hamiltonian is given by

\begin{align}
    \tanh(2\theta_{\bm{k}\sigma}) = \frac{\gamma^{A_1}_{\bm{k}}}{\lambda + \gamma^{B_2}_{\bm{k}} + \sigma \frac{K}{2}\big(m_B - m_A\big)}.
\end{align}
In order to simplify the expressions, we take $m_B = -m_A$, which is consistent with a Néel phase. The diagonalized Hamiltonian now takes the form

\begin{align}
\begin{aligned}
    H^{\text{Néel}}_{\rm{AFMI}} &= E'_0 + \sum_{\substack{\bm{k}\in\diamondsuit\\\sigma}}\omega_{\bm{k}\sigma}\big(\alpha^{\dagger}_{\bm{k}\sigma}\alpha_{\bm{k}\sigma}  + \beta^{\dagger}_{\bm{k}\sigma}\beta_{\bm{k}\sigma}\big),
\end{aligned}
\end{align}

where 
\begin{align}
    \omega_{\bm{k}\sigma} &= \sqrt{\big(\lambda + \gamma^{B_2}_{\bm{k}} - \sigma K m_A \big)^2 - \big( \gamma^{A_1}_{\bm{k}} \big)^2},
\end{align}
and $E'_0 = E_0 + \sum_{\bm{k}\sigma}\omega_{\bm{k}\sigma}$.\\
\indent The mean-field parameters should be determined self-consistently from minimization of the free energy. The free energy per lattice site is given by 

\begin{align}
\begin{aligned}    
    f &= \frac{E'_0}{N} + \frac{2}{\beta N}\sum_{\substack{\bm{k}\in\diamondsuit \\ \sigma}}\ln(1 - e^{-\beta\omega_{\bm{k}\sigma}}).
\end{aligned}
\end{align} 
Minimizing the free energy with respect to $\mathcal{A}_{\bm{\delta}_1}$, $\mathcal{B}_{\bm{\delta}_2}$, $\lambda$ and $m_A$, we then obtain the following self-consistent equations for the mean-field parameters

\begin{subequations}
\begin{align}
    &\mathcal{A}_{\bm{\delta}_1} = \frac{1}{2N}\sum_{\substack{\bm{k}\in\diamondsuit \\ \sigma}}\frac{\gamma^{A_1}_{\bm{k}}}{\omega_{\bm{k}\sigma}}\big(1\! +\! 2n_{\bm{k}\sigma}\big)\cos(\bm{k}\cdot \bm{\delta}_1),\\
    &\mathcal{B}_{\bm{\delta}_2} = \frac{1}{2N}\!\sum_{\substack{\bm{k}\in\diamondsuit \\ \sigma}}\!\frac{\big(\lambda + \gamma^{B_2}_{\bm{k}} - \sigma K m_A \big)}{\omega_{\bm{k}\sigma}} \big(1\! +\! 2n_{\bm{k}\sigma}\big)\cos(\bm{k}\cdot \bm{\delta}_2),\\
    &\bar{\kappa} = \frac{1}{2N}\!\sum_{\substack{\bm{k}\in\diamondsuit \\ \sigma}}\!\frac{\big(\lambda + \gamma^{B_2}_{\bm{k}} - \sigma K m_A \big)}{\omega_{\bm{k}\sigma}}\big(1\! +\! 2n_{\bm{k}\sigma}\big),\\
    & m_A = \frac{1}{2N}\sum_{\substack{\bm{k}\in\diamondsuit \\ \sigma}} \frac{\sigma\big(\lambda + \gamma^{B_2}_{\bm{k}} - \sigma K m_A \big)}{\omega_{\bm{k}\sigma}}\big(1\! +\! 2n_{\bm{k}\sigma}\big).
\end{align}
\end{subequations}
The Bose-Einstein occupation factor is here denoted by $n_{\bm{k}\sigma}$, and we have defined $\bar{\kappa} = \frac{1}{2}\big(\kappa + 1\big)$. As mentioned earlier, in our description of the system, a Néel type state with $m_A > 0$ arises from condensation of $\uparrow$-bosons with $\bm{k} = 0$. For condensation to take place, we need $\abs{\lambda + \gamma^{B_2}_{0} -K m_A} = \abs{\gamma^{A_1}_{0}}$. In the following, we will take $A_{\bm{\delta}_1}$ to be positive. For the interaction potential that will enter into the Hamiltonian describing the effective theory of interacting electrons, we need the ground state properties of the antiferromagnet. From the Bose-Einstein occupation factors, at zero temperature, we then only get a contribution from the condensate, $n_{0\uparrow}$. Defining $\zeta_{\bm{k}\sigma} \equiv \gamma^{A_1}_{0} - \gamma^{B_2}_{0} + \gamma^{B_2}_{\bm{k}} + 2K m_A\delta_{\sigma,\downarrow}$, we now have

\begin{subequations}
\begin{align}
    &\bar{\kappa} = \frac{1}{2N}{\sum_{\substack{\bm{k}\in\diamondsuit \\ \sigma}}}'\,\frac{\zeta_{\bm{k}\sigma}}{\omega_{\bm{k}\sigma}} + \frac{1}{N}Q_0,\\
    &m_A = \frac{1}{2N}{\sum_{\substack{\bm{k}\in\diamondsuit \\ \sigma}}}'\, \frac{\sigma\,\zeta_{\bm{k}\sigma}}{\omega_{\bm{k}\sigma}} + \frac{1}{N}Q_0,\\
    &\mathcal{A}_{\bm{\delta}_1} = \frac{1}{2N}{\sum_{\substack{\bm{k}\in\diamondsuit \\ \sigma}}}'\,\frac{\gamma^{A_1}_{\bm{k}}}{\omega_{\bm{k}\sigma}}\cos(\bm{k}\cdot \bm{\delta}_1) + \frac{1}{N}Q_0,\\
    &\mathcal{B}_{\bm{\delta}_2} =  \frac{1}{2N}{\sum_{\substack{\bm{k}\in\diamondsuit \\ \sigma}}}'\,\frac{\zeta_{\bm{k}\sigma}}{\omega_{\bm{k}\sigma}}\cos(\bm{k}\cdot \bm{\delta}_2) +\frac{1}{N}Q_0,
\end{align}
\end{subequations}

where

\begin{align}
    Q_0 = \frac{\gamma^{A_1}_{0}}{\omega_{0\uparrow}}\Big(n_{0\uparrow}+ \frac{1}{2}\Big).
\end{align}
Note that the sums no longer include $\bm{k} = 0$, $\sigma = \uparrow$. We can then eliminate 

\begin{align}
\begin{aligned}
    \Tilde{Q}_0 \equiv \frac{1}{N}Q_0  = \bar{\kappa} -  \frac{1}{2N}{\sum_{\substack{\bm{k}\in\diamondsuit \\ \sigma}}}'\frac{\zeta_{\bm{k}\sigma}}{\omega_{\bm{k}\sigma}},
\end{aligned}    
\end{align}
and obtain

\begin{subequations}
\begin{align}
    &m_A  - \bar{\kappa} + \frac{1}{N}{\sum_{\bm{k}\in\diamondsuit}}'\frac{\zeta_{\bm{k}\downarrow}}{\omega_{\bm{k}\downarrow}} = 0,\\
    &\mathcal{A} - \bar{\kappa} - \frac{1}{2N}{\sum_{\substack{\bm{k}\in\diamondsuit \\ \sigma}}}'\,\frac{\gamma^{A_1}_{\bm{k}}\cos(k_x) - \zeta_{\bm{k}\sigma}}{\omega_{\bm{k}\sigma}} = 0,\\
    &\mathcal{B} - \bar{\kappa} - \frac{1}{2N}{\sum_{\substack{\bm{k}\in\diamondsuit \\ \sigma}}}'\frac{\zeta_{\bm{k}\sigma}}{\omega_{\bm{k}\sigma}}\Big[\cos(k_x + k_y) - 1\Big] = 0,
\end{align}
\end{subequations}
where we have taken $\mathcal{A} \equiv  \mathcal{A}_{\bm{\delta}_1}$, and $\mathcal{B} \equiv  \mathcal{B}_{\bm{\delta}_2}$. In the thermodynamic limit, we can convert the sums to integrals and solve the coupled set of equations numerically using a 
multidimensional root-finder \cite{GNU}.\\
\indent Solving the self-consistent equations for the mean-field parameters, the properties of the antiferromagnet can be determined e.g.\! for different values of $J_2$. In Fig.\! \ref{fig:Neel_disp_rel}, the Schwinger boson dispersion relation $\omega_{\bm{k}\sigma}$ is presented both deep into the Néel regime and close to the transition to the stripe phase. Note how local minima have developed close to the zone-edges of the Brillouin-zone for $J_2/J_1=0.5$, which are nearly degenerate with the dispersion minimum at the zone center. This indicates the vicinity of an instability of the Néel state into a new spin-ordered state. For the same parameters, the Schwinger boson coherence factor $u_{\bm{k}\sigma}$ is presented in Fig.\! \ref{fig:Neel_u}. All quantities are displayed for $\sigma = \downarrow$, because, as we will see in the following, $\omega_{\bm{k}\downarrow}$ and $u_{\bm{k}\downarrow}$, $v_{\bm{k}\downarrow}$ are the quantities, arising in the effective electron-electron interaction potential, that correspond to the magnon energies and coherence factors encountered in the Holstein-Primakoff treatment of the AFMI.

\begin{figure}[ht] 
    \begin{center}
        \includegraphics[width=1.1\columnwidth,trim= 1.9cm 0.2cm 0.2cm 0.1cm,clip=true]{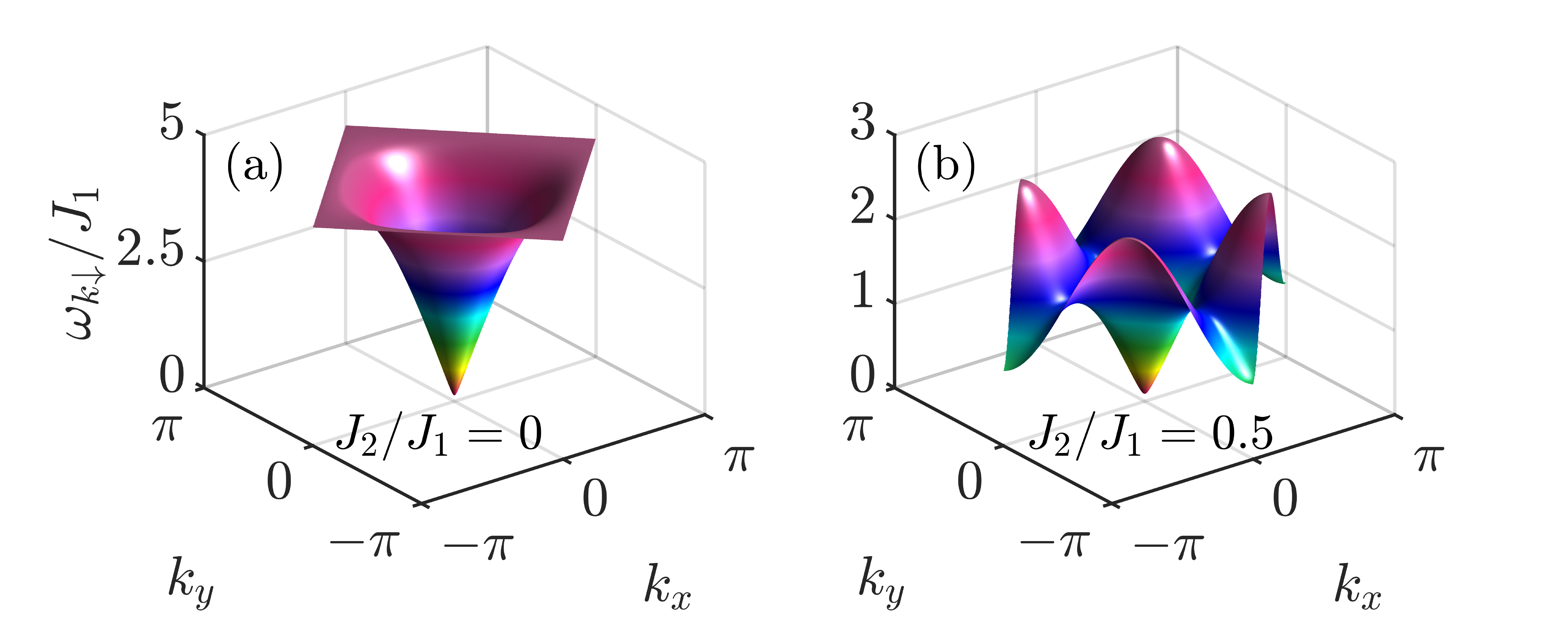}
    \end{center}
    \caption{Schwinger boson dispersion relations $\omega_{\bm{k}\sigma}$ for $K = 0.001 J_1$, $S=1$, and $\sigma = \downarrow$. The value for $J_2$ varies between the two subfigures.}
    \label{fig:Neel_disp_rel}
\end{figure}

\begin{figure}[t] 
    \begin{center}
        \includegraphics[width=1.1\columnwidth,trim= 1.9cm 0.2cm 0.2cm 0.1cm,clip=true]{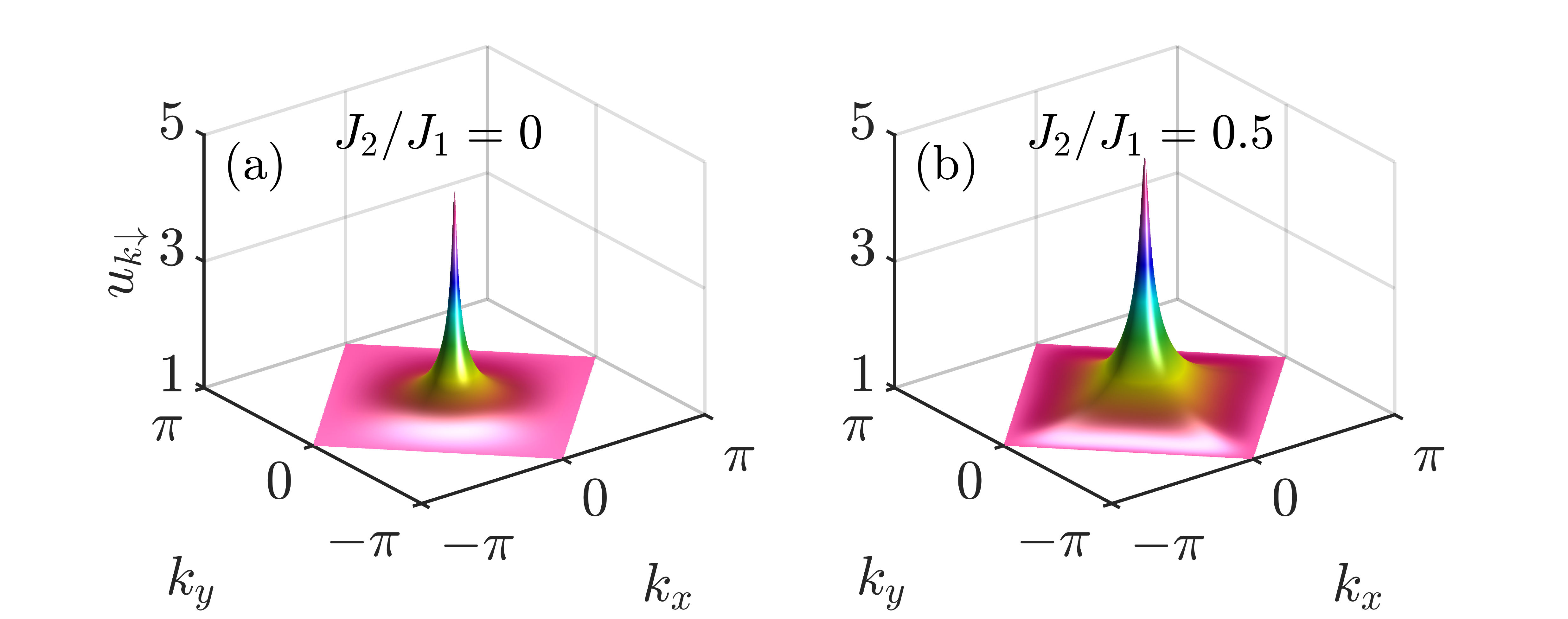} 
    \end{center}
    \caption{Schwinger boson coherence factors $u_{\bm{k}\sigma}$ for $K = 0.001 J_1$, $S=1$, and $\sigma = \downarrow$.}
    \label{fig:Neel_u}
\end{figure}

\subsection{Stripe phase}\label{Secion:Stripe_phase}

For the stripe phase, we assume that the stripes are oriented in the $y$-direction. The results for the superconductivity are not expected to depend on the spatial direction of the stripes. We can then write out the AFMI Hamiltonian as  

\begin{align}
\begin{aligned}
    H^{\text{Stripe}}_{\rm{AFMI}} &= J_1 \!\sum_{\substack{\bm{i} \in A\\ \bm{j} = \bm{i} \pm \hat{x}   }}\!\bm{S}^{A}_{\bm{i}}\cdot \bm{S}^{B}_{\bm{j}} + J_1\! \sum_{\substack{\bm{i} \in A\\ \bm{j} = \bm{i} \pm \hat{y}   }}\!\bm{S}^{A}_{\bm{i}}\cdot \bm{S}^{A}_{\bm{j}}\\
    &+ J_1 \!\sum_{\substack{\bm{i} \in B\\\bm{j} = \bm{i} \pm \hat{x}}}\!\bm{S}^{B}_{\bm{i}}\cdot \bm{S}^{A}_{\bm{j}} + J_1 \!\sum_{\substack{\bm{i} \in B\\\bm{j} = \bm{i} \pm \hat{y}}}\!\bm{S}^{B}_{\bm{i}}\cdot \bm{S}^{B}_{\bm{j}}\\
    &+ J_2\! \sum_{\substack{\bm{i} \in A\\\bm{j} \text{nnn} \bm{i}}}\bm{S}^{A}_{\bm{i}}\cdot \bm{S}^{B}_{\bm{j}} + J_2\! \sum_{\substack{\bm{i} \in B\\\bm{j} \text{nnn} \bm{i}}}\bm{S}^{B}_{\bm{i}}\cdot \bm{S}^{A}_{\bm{j}}\\
    &- K\sum_{\bm{i}\in A}S^2_{\bm{i}z} - K\sum_{\bm{i}\in B}S^2_{\bm{i}z}.
\end{aligned}
\end{align}

The bond operators, this time, take the form 

\begin{subequations}
\begin{align}
    A^{A_x}_{\bm{i}\bm{j}} &= \frac{1}{2}\big(a_{\bm{i}\uparrow}b_{\bm{j}\uparrow} + a_{\bm{i}\downarrow}b_{\bm{j}\downarrow}\big) = A^{2,A}_{\bm{i}\bm{j}},\\
    B^{A_x}_{\bm{i}\bm{j}} &= \frac{1}{2}\big(a_{\bm{i}\downarrow}b^{\dagger}_{\bm{j}\uparrow} - a_{\bm{i}\uparrow}b^{\dagger}_{\bm{j}\downarrow}\big) = B^{2,A}_{\bm{i}\bm{j}},\\
    A^{A_y}_{\bm{i}\bm{j}} &= \frac{1}{2}\big(a_{\bm{i}\uparrow}a_{\bm{j}\downarrow} - a_{\bm{i}\downarrow}a_{\bm{j}\uparrow}\big),\\
    B^{A_y}_{\bm{i}\bm{j}} &= \frac{1}{2}\big(a_{\bm{i}\uparrow}a^{\dagger}_{\bm{j}\uparrow} + a_{\bm{i}\downarrow}a^{\dagger}_{\bm{j}\downarrow}\big),\\
    A^{B_x}_{\bm{i}\bm{j}} &= \frac{1}{2}\big(-b_{\bm{i}\downarrow}a_{\bm{j}\downarrow} - b_{\bm{i}\uparrow}a_{\bm{j}\uparrow}\big) = A^{2,B}_{\bm{i}\bm{j}},\\
    B^{B_x}_{\bm{i}\bm{j}} &= \frac{1}{2}\big(b_{\bm{i}\uparrow}a^{\dagger}_{\bm{j}\downarrow} - b_{\bm{i}\downarrow}a^{\dagger}_{\bm{j}\uparrow}\big) = B^{2,B}_{\bm{i}\bm{j}},\\
    A^{B_y}_{\bm{i}\bm{j}} &= \frac{1}{2}\big(b_{\bm{i}\uparrow}b_{\bm{j}\downarrow} - b_{\bm{i}\downarrow}b_{\bm{j}\uparrow}\big),\\
    B^{B_y}_{\bm{i}\bm{j}} &= \frac{1}{2}\big(b_{\bm{i}\downarrow}b^{\dagger}_{\bm{j}\downarrow} + b_{\bm{i}\uparrow}b^{\dagger}_{\bm{j}\uparrow}\big).
\end{align}
\end{subequations}
Including the Lagrange multiplier terms, the AFMI Hamiltonian can then be written out as 

\begin{align}
\begin{aligned}
    H^{\text{Stripe}}_{\rm{AFMI}} &= J_1 \sum_{\substack{\bm{i} \in A\\ \bm{j} = \bm{i} \pm \hat{x}}}\Big[\big(B^{A_x}_{\bm{i}\bm{j}}\big)^{\dagger}B^{A_x}_{\bm{i}\bm{j}} - \big(A^{A_x}_{\bm{i}\bm{j}}\big)^{\dagger}A^{A_x}_{\bm{i}\bm{j}} - \frac{1}{4}n_{\bm{i},A}\Big]\\
    &+ J_1 \sum_{\substack{\bm{i} \in A\\ \bm{j} = \bm{i} \pm \hat{y}   }}\Big[\big(B^{A_y}_{\bm{i}\bm{j}}\big)^{\dagger}B^{A_y}_{\bm{i}\bm{j}} - \big(A^{A_y}_{\bm{i}\bm{j}}\big)^{\dagger}A^{A_y}_{\bm{i}\bm{j}} - \frac{1}{4}n_{\bm{i},A}\Big]\\
     &+ J_1 \sum_{\substack{\bm{i} \in B\\\bm{j} = \bm{i} \pm \hat{x}}}\Big[\big(B^{B_x}_{\bm{i}\bm{j}}\big)^{\dagger}B^{B_x}_{\bm{i}\bm{j}} - \big(A^{B_x}_{\bm{i}\bm{j}}\big)^{\dagger}A^{B_x}_{\bm{i}\bm{j}} - \frac{1}{4}n_{\bm{i},B}\Big]\\
     &+ J_1 \sum_{\substack{\bm{i} \in B\\\bm{j} = \bm{i} \pm \hat{y}}}\Big[\big(B^{B_y}_{\bm{i}\bm{j}}\big)^{\dagger}B^{B_y}_{\bm{i}\bm{j}} - \big(A^{B_y}_{\bm{i}\bm{j}}\big)^{\dagger}A^{B_y}_{\bm{i}\bm{j}} - \frac{1}{4}n_{\bm{i},B}\Big]\\
    &+ J_2\! \sum_{\substack{\bm{i} \in A\\\bm{j} \text{nnn} \bm{i}}}\Big[\big(B^{2,A}_{\bm{i}\bm{j}}\big)^{\dagger}B^{2,A}_{\bm{i}\bm{j}} - \big(A^{2,A}_{\bm{i}\bm{j}}\big)^{\dagger}A^{2,A}_{\bm{i}\bm{j}} - \frac{1}{4}n_{\bm{i},A}\Big]\\
    &+ J_2\! \sum_{\substack{\bm{i} \in B\\\bm{j} \text{nnn} \bm{i}}}\Big[\big(B^{2,B}_{\bm{i}\bm{j}}\big)^{\dagger}B^{2,B}_{\bm{i}\bm{j}} - \big(A^{2,B}_{\bm{i}\bm{j}}\big)^{\dagger}A^{2,B}_{\bm{i}\bm{j}} - \frac{1}{4}n_{\bm{i},B}\Big]\\
    &- K\sum_{\bm{i}\in A}S^2_{\bm{i}z} - K\sum_{\bm{i}\in B}S^2_{\bm{i}z} + \lambda_A\sum_{\bm{i} \in A}\big(n_{\bm{i},A} - \kappa\big)\\
    &+ \lambda_B\sum_{\bm{i} \in B}\big(n_{\bm{i},B} - \kappa\big).
\end{aligned}
\end{align}
\indent In order to make progress, we introduce the mean-field decoupling from Eq.\! \eqref{eq:C}, and choose an Ansatz for the mean-field parameters that will produce a stripe phase

\begin{subequations}
\begin{align*}
    &\langle A^{A_x}_{\bm{i}\bm{j}} \rangle  = -\langle A^{B_x}_{\bm{i}\bm{j}} \rangle \equiv \mathcal{A}_{\bm{\delta}_x},\\
    &\langle B^{A_x}_{\bm{i}\bm{j}} \rangle  = \langle B^{B_x}_{\bm{i}\bm{j}} \rangle = 0,\\
    &\langle A^{A_y}_{\bm{i}\bm{j}} \rangle  = \langle A^{B_y}_{\bm{i}\bm{j}} \rangle = 0,\\
    &\langle B^{A_y}_{\bm{i}\bm{j}} \rangle  = \langle B^{B_y}_{\bm{i}\bm{j}} \rangle \equiv \mathcal{B}_{\bm{\delta}_y},\\
    &\langle B^{2,A}_{\bm{i}\bm{j}} \rangle  = \langle B^{2,B}_{\bm{i}\bm{j}} \rangle = 0,\\
    &\langle A^{2,A}_{\bm{i}\bm{j}} \rangle  = -\langle A^{2,B}_{\bm{i}\bm{j}} \rangle \equiv \mathcal{A}_{\bm{\delta}_2}.
\end{align*}
\end{subequations}

\begin{figure}[ht] 
    \begin{center}
        \includegraphics[width=0.95\columnwidth,trim= 3.2cm 21.5cm 5.0cm 0.7cm,clip=true]{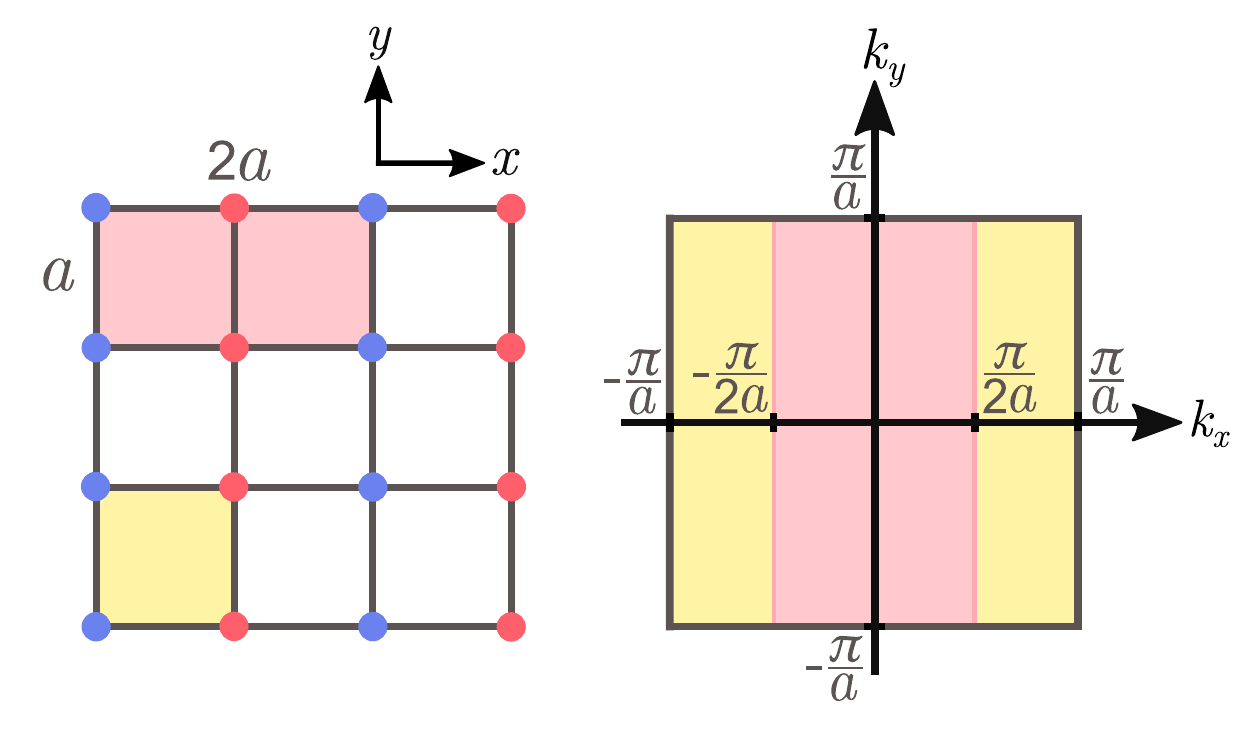}
    \end{center}
    \caption{Unit cell and Brillouin zone for the full lattice and the sublattices.}
    \label{fig:Stripe_k}
\end{figure}
\begin{figure}[ht] 
    \begin{center}
        \includegraphics[width=1.1\columnwidth,trim= 1.9cm 0.2cm 0.2cm 0.4cm,clip=true]{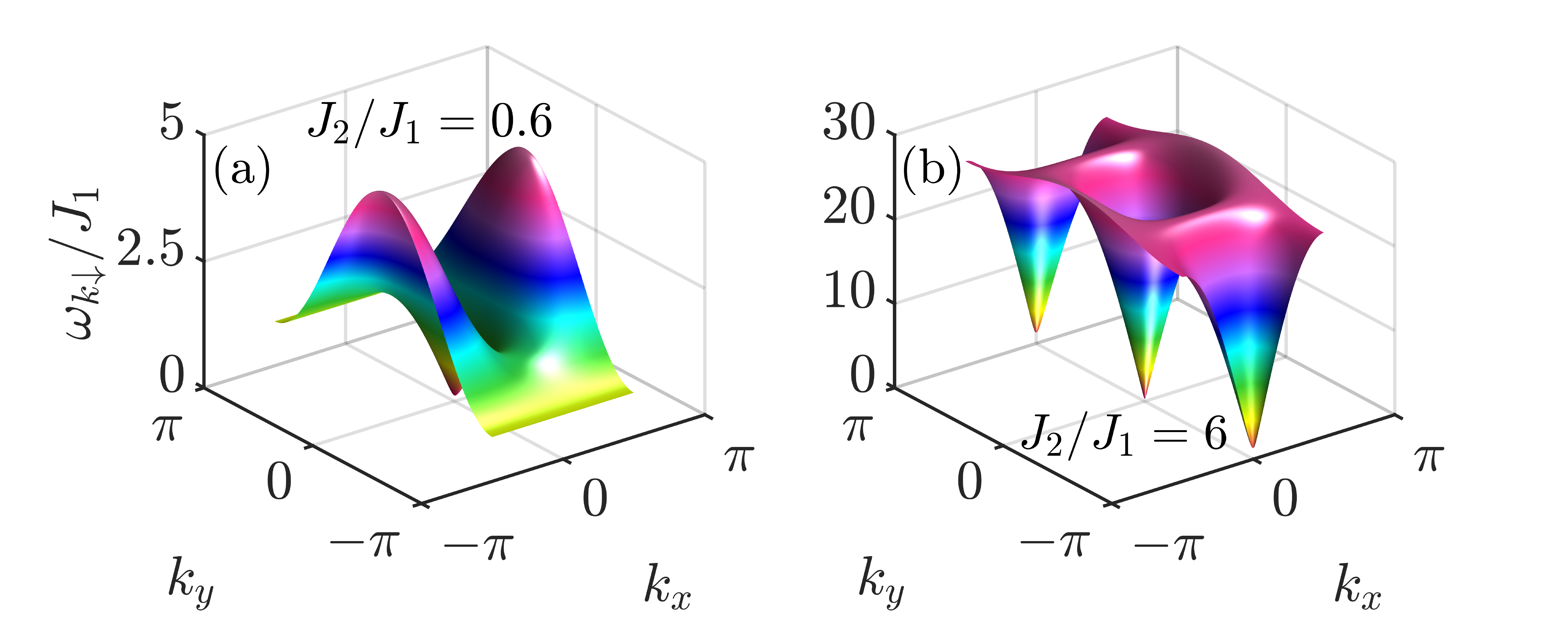}
    \end{center}
    \caption{Schwinger boson dispersion relations $\omega_{\bm{k}\sigma}$ for $K = 0.001 J_1$, $S=1$, and $\sigma = \downarrow$. The value for $J_2$ varies between the two subfigures.}
    \label{fig:Stripe_disp_rel}
\end{figure}

\begin{figure}[ht] 
    \begin{center}
        \includegraphics[width=1.1\columnwidth,trim= 1.9cm 0.2cm 0.2cm 0.4cm,clip=true]{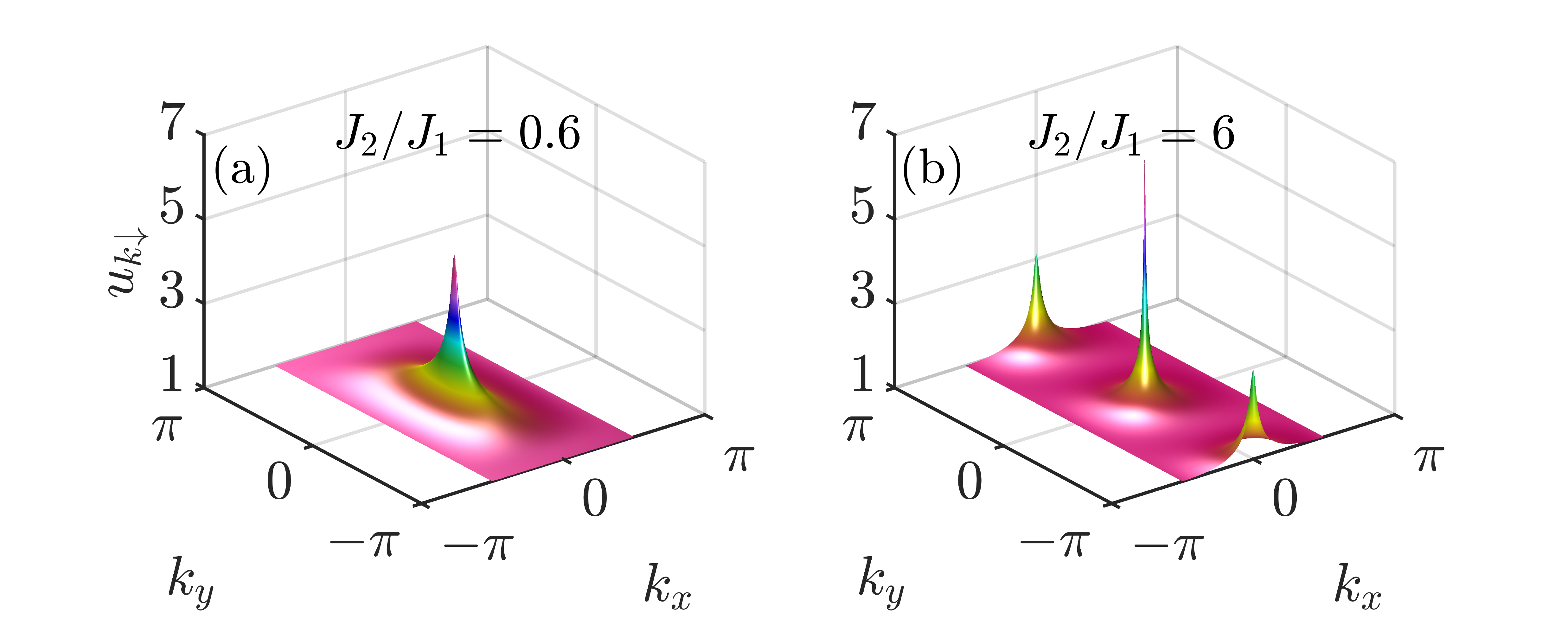}
    \end{center}
    \caption{Schwinger boson coherence factors $u_{\bm{k}\sigma}$ for $K = 0.001 J_1$, $S=1$, and $\sigma = \downarrow$.}
    \label{fig:Stripe_u}
\end{figure}

\begin{figure}[ht] 
    \begin{center}
        \includegraphics[width=0.8\columnwidth,trim= 2.7cm 7.7cm 2.0cm 8.5cm,clip=true]{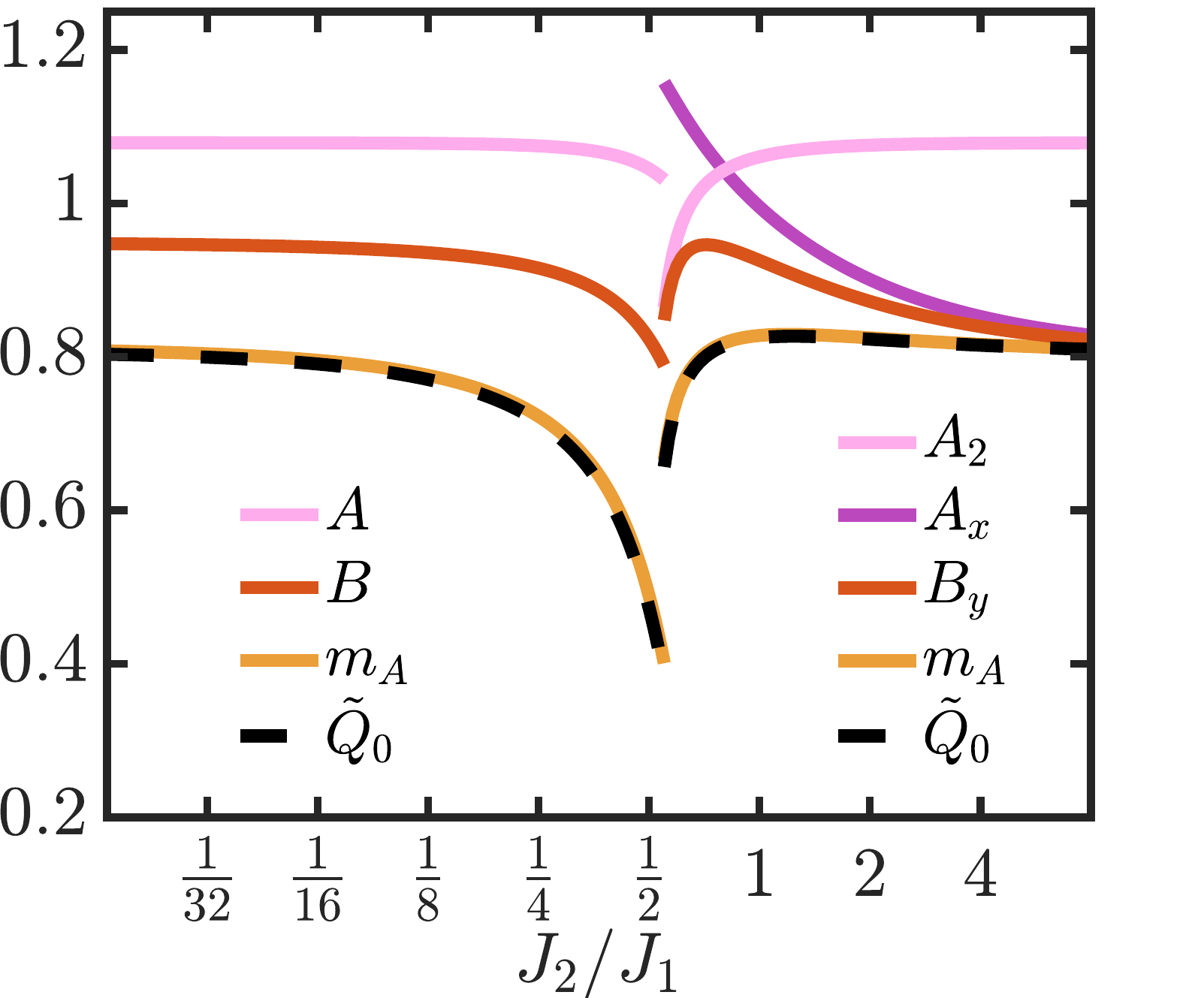}
    \end{center}
    \caption{Mean-field parameters for the Néel and stripe phase for $K/J_1 = 0.001$ and $S = 1$. The transition between the two phases is found to take place around $J_2/J_1 = 0.549$.}
    \label{fig:MF_params}
\end{figure}

 
\noindent As before, we take $\lambda_A = \lambda_B \equiv \lambda$ and treat the easy-axis anisotropy terms in the same mean-field fashion as for the Néel state. Next, we introduce Fourier transformations for the boson operators

\begin{subequations}
\begin{align}
    &a_{\bm{i}\sigma} = \frac{1}{\sqrt{N_A}}\sum_{\bm{k}\in\vrec}e^{i\bm{k}\cdot \bm{r}_{\bm{i}}}a_{\bm{k}\sigma},\\
    &b_{\bm{i}\sigma} = \frac{1}{\sqrt{N_B}}\sum_{\bm{k}\in\vrec}e^{i\bm{k}\cdot \bm{r}_{\bm{i}}}b_{\bm{k}\sigma},
\end{align}
\end{subequations}
where the sum over momentum covers the reduced Brillouin zone of the sublattices $\vrec$, as displayed in Fig.\! \ref{fig:Stripe_k}. The Hamiltonian then takes the form

\begin{align}
\begin{aligned}
    &H^{\text{Stripe}}_{\rm{AFMI}} = 2 N_A \Big[J_1\sum_{\bm{\delta}_x}\big(\mathcal{A}_{\bm{\delta}_x}\big)^2 - J_1\sum_{\bm{\delta}_y}\big(\mathcal{B}_{\bm{\delta}_y}\big)^2\\
    &+ J_2\sum_{\bm{\delta}_2}\big(\mathcal{A}_{\bm{\delta}_2}\big)^2 \Big] - 2N_A\kappa \lambda + K N_A( m^2_A + m^2_B) \\
    &+ \Big[\lambda - \frac{1}{4}(J_1 z_1 + J_2 z_2)\Big]\sum_{\substack{\bm{k}\in \vrec\\\sigma}}\big(a^{\dagger}_{\bm{k}\sigma}a_{\bm{k}\sigma} + b^{\dagger}_{\bm{k}\sigma}b_{\bm{k}\sigma}\big)\\
    &- K m_A\! \sum_{\substack{\bm{k}\in\vrec \\ \sigma}}\sigma\, a^{\dagger}_{\bm{k}\sigma}a_{\bm{k}\sigma} + K m_B\!\sum_{\substack{\bm{k}\in\vrec \\ \sigma}}\sigma\, b^{\dagger}_{\bm{k}\sigma}b_{\bm{k}\sigma}\\
    &+ \sum_{\substack{\bm{k}\in \vrec\\\sigma}}\gamma^{B_y}_{\bm{k}}\Big(a^{\dagger}_{\bm{k}\sigma}a_{\bm{k}\sigma} + b_{\bm{k}\sigma}b^{\dagger}_{\bm{k}\sigma}\Big)\\
    &- \sum_{\substack{\bm{k}\in\vrec\\ \sigma}} \Big(\gamma^{A_x}_{\bm{k}} + \gamma^{A_2}_{\bm{k}}\Big)\Big( b^{\dagger}_{\bm{k}\sigma}a^{\dagger}_{-\bm{k}\sigma} + a_{\bm{k}\sigma}b_{-\bm{k}\sigma}\Big),
\end{aligned}
\end{align}

where we have defined

\begin{subequations}
\begin{align}
    &\gamma^{A_x}_{\bm{k}} = J_1\! \sum_{\bm{\delta}_x}\mathcal{A}_{\bm{
    \delta}_x}  \cos(\bm{k}\cdot \bm{\delta_x}),\\
    &\gamma^{B_y}_{\bm{k}} = J_1\! \sum_{\bm{\delta}_y}\mathcal{B}_{\bm{
    \delta}_y}  \cos(\bm{k}\cdot \bm{\delta_y}),\\
    &\gamma^{A_2}_{\bm{k}} = J_2\! \sum_{\bm{\delta}_2}\mathcal{A}_{\bm{
    \delta}_2}  \cos(\bm{k}\cdot \bm{\delta_2}).
\end{align}
\end{subequations}
The sums over nearest neighbors still cover both positive and negative directions. We then redefine $\lambda$ as we did for the Néel phase and exclude constant terms not involving mean-field parameters. Splitting up the Hamiltonian in three parts $H^{\text{Stripe}}_{\rm{AFMI}} = E_0 + H^{\uparrow} + H^{\downarrow}$, we write

\begin{align}
\begin{aligned}
    E_0 &= 2 N_A \Big[J_1\sum_{\bm{\delta}_x}\big(\mathcal{A}_{\bm{\delta}_x}\big)^2 - J_1\sum_{\bm{\delta}_y}\big(\mathcal{B}_{\bm{\delta}_y}\big)^2 + J_2\sum_{\bm{\delta}_2}\big(\mathcal{A}_{\bm{\delta}_2}\big)^2 \Big]\\
    &- 2N_A\lambda(\kappa + 1) + K N_A( m^2_A + m^2_B),
\end{aligned}
\end{align}

\begin{align}
\begin{aligned}
    H^{\sigma} &= \sum_{\bm{k}\in\vrec}\big(\lambda + \gamma^{B_y}_{\bm{k}} - \sigma K m_A\big) a^{\dagger}_{\bm{k}\sigma}a_{\bm{k}\sigma}\\
    &+ \sum_{\bm{k}\in\vrec}\big(\lambda + \gamma^{B_y}_{\bm{k}} + \sigma K m_B\big) b_{\bm{k}\sigma}b^{\dagger}_{\bm{k}\sigma}\\
    &-\sum_{\bm{k}\in\vrec} \big(\gamma^{A_x}_{\bm{k}} + \gamma^{A_2}_{\bm{k}}\big)\big( b^{\dagger}_{\bm{k}\sigma}a^{\dagger}_{-\bm{k}\sigma} +  a_{\bm{k}\sigma}b_{-\bm{k}\sigma}\big).
\end{aligned}
\end{align}
As in the Néel case, the Bogoliubov transformation of Eq.\! \eqref{eq:bogoliubov} diagonalizes the Hamiltonian, where $\theta_{\bm{k}\sigma}$ this time is given by

\begin{align}
    \tanh(2\theta_{\bm{k}\sigma}) = \frac{\gamma^{A_x}_{\bm{k}} + \gamma^{A_2}_{\bm{k}}}{\lambda + \gamma^{B_y}_{\bm{k}} + \sigma\frac{K}{2}\big(m_B - m_A\big)}.
\end{align}
Taking $m_B = - m_A$, we obtain the following expression for the diagonalized Hamiltonian

\begin{align}
\begin{aligned}
    H^{\text{Stripe}}_{\rm{AFMI}} &= E'_0 + \sum_{\substack{\bm{k}\in\vrec\\\sigma}}\omega_{\bm{k}\sigma}\big(\alpha^{\dagger}_{\bm{k}\sigma}\alpha_{\bm{k}\sigma}  + \beta^{\dagger}_{\bm{k}\sigma}\beta_{\bm{k}\sigma}\big),
\end{aligned}
\end{align}
where 
\begin{align}
\begin{aligned}
    \omega_{\bm{k}\sigma} = \sqrt{\big(\lambda + \gamma^{B_y}_{\bm{k}} - \sigma K m_A \big)^2 - \big(\gamma^{A_x}_{\bm{k}} + \gamma^{A_2}_{\bm{k}}\big)^2},
\end{aligned}
\end{align}
and $E'_0 = E_0 + \sum_{\bm{k}\sigma}\omega_{\bm{k}\sigma}$. Minimizing the free energy with respect to $\mathcal{A}_{\bm{\delta}_x}$, $\mathcal{A}_{\bm{\delta}_2}$, $\mathcal{B}_{\bm{\delta}_y}$, $\lambda$, and $m_A$, we obtain 

\begin{subequations}
\begin{align}
    &\mathcal{A}_{\bm{\delta}_x}  = \frac{1}{2N}\sum_{\substack{\bm{k}\in\vrec \\ \sigma}}\frac{\big(\gamma^{A_x}_{\bm{k}} + \gamma^{A_2}_{\bm{k}}\big)}{\omega_{\bm{k}\sigma}}\big(1 + 2n_{\bm{k}\sigma}\big) \cos(\bm{k}\cdot \bm{\delta}_x),\\
    &\mathcal{A}_{\bm{\delta}_2}  = \frac{1}{2N}\sum_{\substack{\bm{k}\in\vrec \\ \sigma}}\frac{\big(\gamma^{A_x}_{\bm{k}} + \gamma^{A_2}_{\bm{k}}\big)}{\omega_{\bm{k}\sigma}}\big(1 + 2n_{\bm{k}\sigma}\big) \cos(\bm{k}\cdot \bm{\delta}_2),\\
    &\mathcal{B}_{\bm{\delta}_y}  = \frac{1}{2N}\sum_{\substack{\bm{k}\in\vrec \\ \sigma}}\frac{\big(\lambda + \gamma^{B_y}_{\bm{k}} - \sigma K m_A \big)}{\omega_{\bm{k}\sigma}}\big(1 + 2n_{\bm{k}\sigma}\big) \cos(\bm{k}\cdot \bm{\delta}_y),\\
    &\bar{\kappa}  = \frac{1}{2N}\sum_{\substack{\bm{k}\in\vrec \\ \sigma}}\frac{\big(\lambda + \gamma^{B_y}_{\bm{k}} - \sigma K m_A \big)}{\omega_{\bm{k}\sigma}}\big(1 + 2n_{\bm{k}\sigma}\big),\\
    &m_A = \frac{1}{2N}\sum_{\substack{\bm{k}\in\vrec \\ \sigma}}\frac{\sigma\big(\lambda + \gamma^{B_y}_{\bm{k}} - \sigma K m_A \big)}{\omega_{\bm{k}\sigma}}\big(1 + 2n_{\bm{k}\sigma}\big).
\end{align}
\end{subequations}
Here, we have once again defined $\bar{\kappa} = \frac{1}{2}\big(\kappa + 1\big)$. For a stripe-type state with $m_A > 0$, we will have condensation of $\uparrow$-bosons with $\bm{k} = 0$, and we then need $\abs{\lambda + \gamma^{B_y}_{0} - K m_A} = \abs{\gamma^{A_x}_{0} + \gamma^{A_2}_{0}}$. The mean-field parameters $\mathcal{A}_{\bm{\delta}_x}$ and $\mathcal{A}_{\bm{\delta}_2}$ will be assumed to be positive. At zero temperature, we once again only get contributions from $n_{\bm{k}\sigma}$ from the condensate. We then define $\zeta_{\bm{k}\sigma} = \gamma^{A_x}_{0} + \gamma^{A_2}_{0} - \gamma^{B_y}_{0} + \gamma^{B_y}_{\bm{k}} + 2 K m_A \delta_{\sigma,\downarrow}$, and write

\begin{subequations}
\begin{align}
    &\bar{\kappa}  = \frac{1}{2N}{\sum_{\substack{\bm{k}\in\vrec \\ \sigma}}}'\frac{\zeta_{\bm{k}\sigma}}{\omega_{\bm{k}\sigma}}+ \frac{1}{N}Q_0,\\
    &m_A = \frac{1}{2N}{\sum_{\substack{\bm{k}\in\vrec \\ \sigma}}}'\frac{\sigma\,\zeta_{\bm{k}\sigma}}{\omega_{\bm{k}\sigma}}+ \frac{1}{N}Q_0,\\
    &\mathcal{A}_{x}  = \frac{1}{2N}{\sum_{\substack{\bm{k}\in\vrec \\ \sigma}}}'\frac{\big(\gamma^{A_x}_{\bm{k}} + \gamma^{A_2}_{\bm{k}}\big)}{\omega_{\bm{k}\sigma}} \cos(k_x) + \frac{1}{N}Q_0,\\
    &\mathcal{A}_{2}  = \frac{1}{2N}{\sum_{\substack{\bm{k}\in\vrec \\ \sigma}}}'\frac{\big(\gamma^{A_x}_{\bm{k}} + \gamma^{A_2}_{\bm{k}}\big)}{\omega_{\bm{k}\sigma}} \cos(k_x + k_y) + \frac{1}{N}Q_0,\\
    &\mathcal{B}_{y}  = \frac{1}{2N}{\sum_{\substack{\bm{k}\in\vrec \\ \sigma}}}'\frac{\zeta_{\bm{k}\sigma}}{\omega_{\bm{k}\sigma}} \cos(k_y) + \frac{1}{N}Q_0,
\end{align}
\end{subequations}

where 

\begin{align}
    Q_0 = \frac{\big(\gamma^{A_x}_{0} + \gamma^{A_2}_{0}\big)}{\omega_{0\uparrow}}\Big(n_{0\uparrow}+ \frac{1}{2}\Big),
\end{align}
and we have taken $\mathcal{A}_x \equiv \mathcal{A}_{\bm{\delta}_x}$, $\mathcal{A}_2 \equiv \mathcal{A}_{\bm{\delta}_2}$, and $\mathcal{B}_y \equiv \mathcal{B}_{\bm{\delta}_y}$. We can then once again eliminate $Q_0$

\begin{align}
\begin{aligned}
    \Tilde{Q}_0 \equiv \frac{1}{N}Q_0 = \bar{\kappa}  - \frac{1}{2N}{\sum_{\bm{k}\sigma}}'\frac{\zeta_{\bm{k}\sigma}}{\omega_{\bm{k}\sigma}},
\end{aligned}    
\end{align}

and obtain the equations

\begin{subequations}
\begin{align}
    &m_A = \bar{\kappa} - \frac{1}{N}{\sum_{\bm{k}\in\vrec}}'\frac{\zeta_{\bm{k}\downarrow}}{\omega_{\bm{k}\downarrow}},\\
    &\mathcal{A}_{x} = \bar{\kappa} +  \frac{1}{2N}{\sum_{\substack{\bm{k}\in\vrec \\ \sigma}}}'\frac{\big(\gamma^{A_x}_{\bm{k}} + \gamma^{A_2}_{\bm{k}}\big)\cos(k_x) - \zeta_{\bm{k}\sigma}}{\omega_{\bm{k}\sigma}},\\
    &\mathcal{A}_{2} = \bar{\kappa} + \frac{1}{2N}{\sum_{\substack{\bm{k}\in\vrec \\ \sigma}}}'\frac{\big(\gamma^{A_x}_{\bm{k}} + \gamma^{A_2}_{\bm{k}}\big)\cos(k_x + k_y) - \zeta_{\bm{k}\sigma}}{\omega_{\bm{k}\sigma}},\\
    &\mathcal{B}_{y} = \bar{\kappa} + \frac{1}{2N}{\sum_{\substack{\bm{k}\in\vrec \\ \sigma}}}'\frac{\zeta_{\bm{k}\sigma}}{\omega_{\bm{k}\sigma}} \Big[\cos(k_y) - 1\Big].
\end{align}
\end{subequations}
In the thermodynamic limit, we can then convert the sums into integrals and solve the coupled set of equations numerically.\\
\indent As for the Néel phase, we present dispersion relations and coherence factors for values of $J_2/J_1$ deep into the stripe phase and close to the transition to the other magnetic phase. These results are displayed in Fig.\! \ref{fig:Stripe_disp_rel} and \ref{fig:Stripe_u}. In addition, we also explore in Fig.\! \ref{fig:MF_params} how the mean-field parameters depend on $J_2/J_1$ for the two phases. Notably, the factor $\tilde{Q}_0$, which is closely related to the sublattice magnetization, decreases towards the phase transition and is reduced more on the Néel side of the transition than on the stripe side. This factor will show up later in the effective electron-electron interaction potential. 

\section{Coupling to the NM}\label{Section:Coupling}

\begin{figure}[t] 
    \begin{center}
        \includegraphics[width=0.8\columnwidth,trim= 3.5cm 21.5cm 5.1cm 0.7cm,clip=true]{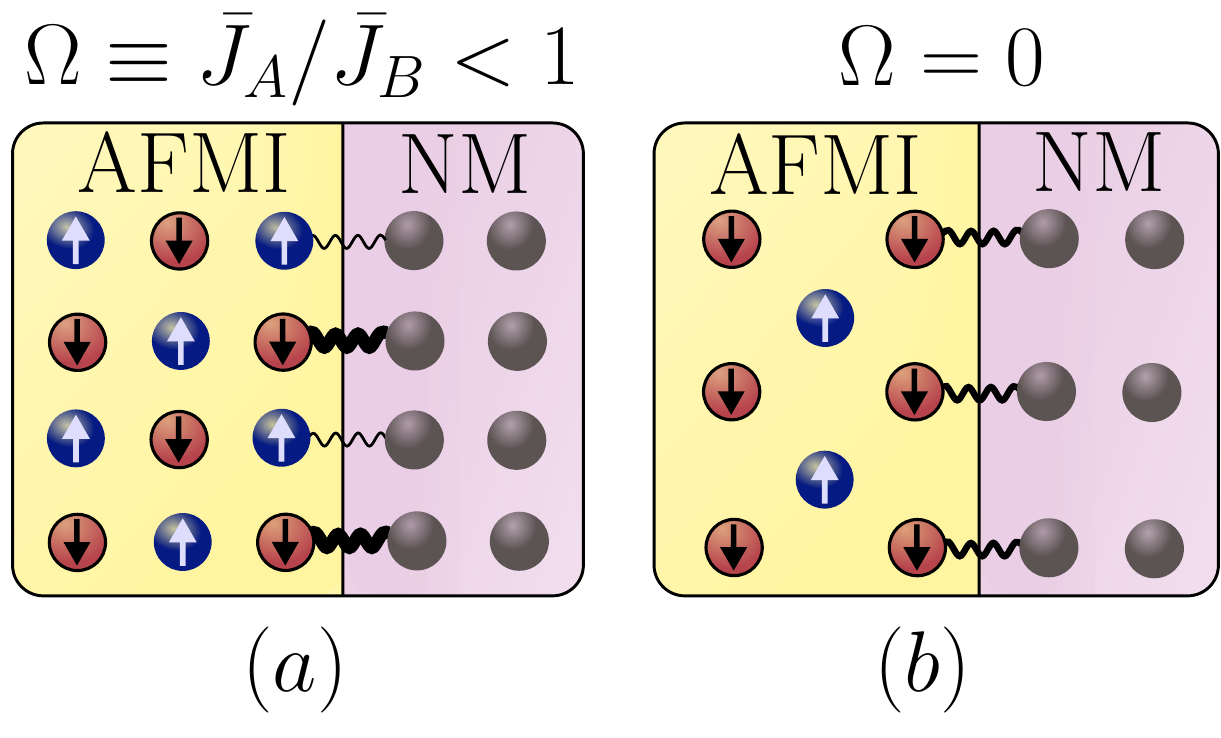}
    \end{center}
    \caption{Panel (a) illustrates our model where the coupling to the two sublattices is allowed to differ. The coupling asymmetry is parametrized by $\Omega$. Panel (b) shows an uncompensated interface, where only one of the two sublattices is present, producing a coupling corresponding to $\Omega = 0$.}
    \label{fig:coupling}
\end{figure}

When considering the coupling to the normal metal (NM), we will treat both the Néel phase and stripe phase simultaneously as the calculation is identical for both phases. It should then be understood that the expressions for the factor $\Tilde{Q}_0$, the Schwinger boson energies $\omega_{\bm{k}\sigma}$, and the Schwinger boson coherence factors $u_{\bm{k}\sigma}$, $v_{\bm{k}\sigma}$, depend on the magnetic phase.\\
\indent Through a Fourier transformation, the NM Hamiltonian is brought on diagonal form

\begin{align}
    H_{\text{NM}} = \sum_{\substack{\bm{k}\in\Box \\ \sigma}}\epsilon_{\bm{k}}c^{\dagger}_{\bm{k}\sigma}c_{\bm{k}\sigma},
\end{align}
with

\begin{align}
    \epsilon_{\bm{k}} = -tz\gamma_{\bm{k}} - \mu.
\end{align}
Here, the sum over momentum covers the Brillouin zone of the full lattice $\Box$. In addition, the NM is exchange-coupled to the two sublattices of the antiferromagnet, $H_{\text{int}} = H^{(A)}_{\text{int}} + H^{(B)}_{\text{int}}$

\begin{align}
\begin{aligned}
    H^{(A)}_{\text{int}} &= -2\Bar{J}\,\Omega\,\sum_{\bm{i}\in A}(a^{\dagger}_{\bm{i}\uparrow} a_{\bm{i}\downarrow}\,c_{\bm{i}\downarrow}^{\dagger}c_{\bm{i}\uparrow} + a^{\dagger}_{\bm{i}\downarrow} a_{\bm{i}\uparrow}\,c_{\bm{i}\uparrow}^{\dagger}c_{\bm{i}\downarrow})\\
    &- \Bar{J}\,\Omega\sum_{\substack{\bm{i}\in A\\\sigma}}\sigma c_{\bm{i}\sigma}^{\dagger}c_{\bm{i}\sigma}(a^{\dagger}_{\bm{i}\uparrow}a_{\bm{i}\uparrow} - a^{\dagger}_{\bm{i}\downarrow}a_{\bm{i}\downarrow}),
\end{aligned}
\end{align}

\begin{align}
\begin{aligned}
    H^{(B)}_{\text{int}} &= 2\Bar{J}\,\sum_{\bm{i}\in B}(b^{\dagger}_{\bm{i}\downarrow} b_{\bm{i}\uparrow}\,c_{\bm{i}\downarrow}^{\dagger}c_{\bm{i}\uparrow} + b^{\dagger}_{\bm{i}\uparrow} b_{\bm{i}\downarrow}\,c_{\bm{i}\uparrow}^{\dagger}c_{\bm{i}\downarrow})\\
    &+ \Bar{J}\,\sum_{\substack{\bm{i}\in B\\\sigma}}\sigma c_{\bm{i}\sigma}^{\dagger}c_{\bm{i}\sigma}(b^{\dagger}_{\bm{i}\uparrow}b_{\bm{i}\uparrow} -  b^{\dagger}_{\bm{i}\downarrow}b_{\bm{i}\downarrow}).
\end{aligned}
\end{align}
Here, we have defined $\Omega \equiv \bar{J}_A/\bar{J}_B$, as visualized in Fig.\! \ref{fig:coupling}, and $\bar{J} \equiv \bar{J}_B$. For magnetic ordering in the z-direction in spin-space, the $z$-component of the coupling gives rise to a staggered magnetic exchange field. For asymmetric coupling to the two sublattices, the NM electrons are then exposed to a net magnetic field, which will influence the superconductivity. For an in-plane magnetic field, the dominant effect is the Pauli pair-breaking mechanism, rather than the orbital effect \cite{Bergeret2018}. As described in Ref.\! \cite{Erlandsen2019}, this paramagnetic effect, arising from the $z$-component of the coupling, is not expected to destroy the superconductivity in the considered system, and can be counteracted by e.g.\! applying an oppositely directed external magnetic field \cite{Lange2003}. The effect of the $z$-component of the coupling will therefore be neglected in the following.  

We then perform Fourier transformations, where the electron operators are transformed as 

\begin{align}
    c_{\bm{i}\sigma} = \frac{1}{\sqrt{N}} \sum_{\bm{k}\in \rm{RBZ}}\Big(c_{\bm{k}\sigma}e^{i\bm{k}\cdot\bm{r}_{\bm{i}}} + c_{\bm{k}+\bm{G},\sigma}e^{i(\bm{k}+ \bm{G})\cdot\bm{r}_{\bm{i}}}\Big),
\end{align}
where $\bm{G} \equiv \frac{\pi(\hat{x} + \hat{y})}{a}$ for the Néel phase and $\bm{G} \equiv \frac{\pi\hat{x}}{a}$ for the stripe phase. The sum over momentum covers the reduced Brillouin zone of the sublattices, $\rm{RBZ}$. Umklapp processes where the momentum of the outgoing electron is shifted by a reciprocal lattice vector of the sublattices, will arise due to the electrons and Schwinger bosons living in different Brillouin zones \cite{Erlandsen2019, Erlandsen2020}. These processes are expected to be important for induced superconductivity in the case of an AFMI coupled to a NM close to half-filling \cite{Fjaerbu2019}. Away from half-filling, the Umklapp processes are, however, of less importance. In addition, for a real uncompensated interface, the NM will be lattice matched with the AFMI sublattice it is coupled to, removing the Umklapp processes from the coupling. The Umklapp processes will therefore not be included in our treatment of the system.\\
\indent We are now left with the coupling terms  

\begin{align}
    H^{(A)}_{\text{int}} &= -\frac{2\Bar{J}\,\Omega}{N}\,\sum_{\bm{q}\in \Box}\sum_{\substack{\bm{k}\bm{k}'\\ \in \rm{RBZ}}}(a^{\dagger}_{\bm{k}\uparrow}a_{\bm{k}'\downarrow}c^{\dagger}_{\bm{q}\downarrow}c_{\bm{k} + \bm{q} - \bm{k}'\uparrow} +  \text{h.c.}),\\
    H^{(B)}_{\text{int}} &= \frac{2\Bar{J}}{N}\,\sum_{\bm{q}\in \Box}\sum_{\substack{\bm{k}\bm{k}'\\ \in \rm{RBZ}}}(b^{\dagger}_{\bm{k}\downarrow}b_{\bm{k}'\uparrow}c^{\dagger}_{\bm{q}\downarrow}c_{\bm{k} + \bm{q} - \bm{k}'\uparrow} + \text{h.c.}).
\end{align}
Expressing the the sublattice Schwinger boson operators in terms of the boson operators that diagonalized the AFMI Hamiltonian, we have the final expression for the electron-boson coupling

\begin{align}
\begin{aligned}
    H^{(A)}_{\text{int}} &= -\frac{2\Bar{J}\,\Omega}{N}\,\sum_{\bm{q}\in \Box}\sum_{\substack{\bm{k}\bm{k}'\\ \in \rm{RBZ}}}\Big[\big(u_{\bm{k}\uparrow} \alpha^{\dagger}_{\bm{k}\uparrow} - v_{\bm{k}\uparrow} \beta_{-\bm{k}\uparrow}\big)\\
    &\times\big(u_{\bm{k}'\downarrow} \alpha_{\bm{k}'\downarrow} - v_{\bm{k}'\downarrow} \beta^{\dagger}_{-\bm{k}'\downarrow}\big)c^{\dagger}_{\bm{q}\downarrow}c_{\bm{k} + \bm{q} - \bm{k}'\uparrow} + \text{h.c.}\Big],
\end{aligned}
\end{align}

\begin{align}
\begin{aligned}
    H^{(B)}_{\text{int}} &= \frac{2\Bar{J}}{N}\,\sum_{\bm{q}\in \Box}\sum_{\substack{\bm{k}\bm{k}'\\ \in \rm{RBZ}}}\Big[\big(v_{\bm{k}\downarrow} \alpha_{-\bm{k}\downarrow} - u_{\bm{k}\downarrow}\beta^{\dagger}_{\bm{k}\downarrow}\big)\\
    &\times\big(v_{\bm{k}'\uparrow} \alpha^{\dagger}_{-\bm{k}'\uparrow} - u_{\bm{k}'\uparrow}\beta_{\bm{k}'\uparrow}\big)c^{\dagger}_{\bm{q}\downarrow}c_{\bm{k} + \bm{q} - \bm{k}'\uparrow} + \text{h.c.}\Big].
\end{aligned}
\end{align}
In the next section, we will derive effective electron-electron interactions, mediated by Schwinger bosons, arising from this electron-boson coupling.

\section{Effective interaction}\label{Section:Eff_int}

We now perform a Schrieffer-Wolff transformation \cite{Schrieffer1966} in order to integrate out the boson operators and obtain an effective theory of interacting electrons. As we have two boson operators in the initial electron-boson coupling, we will end up with electron-electron scattering processes where there are still two boson operators present. The remaining pair of boson operators will be replaced by its ground state expectation value in order to obtain an effective theory of interacting electrons. We define $H = H_0 + \eta H_1$ with

\begin{align}
\begin{aligned}
    H_0 &\equiv E_0' + \sum_{\substack{\bm{k}\in \text{RBZ}\\ \sigma}}\!\omega_{\bm{k}\sigma}\Big(\alpha^{\dagger}_{\bm{k}\sigma}\alpha_{\bm{k}\sigma}  + \beta^{\dagger}_{\bm{k}\sigma}\beta_{\bm{k}\sigma}\Big)\\
    &+ \sum_{\substack{\bm{k}\in\Box \\ \sigma}}\epsilon_{\bm{k}}c^{\dagger}_{\bm{k}\sigma}c_{\bm{k}\sigma},
\end{aligned}
\end{align}
and 

\begin{align}
    \eta H_1 = \eta H_1^{(A)} + \eta H_1^{(B)} \equiv H_{\text{int}}^{(A)} + H_{\text{int}}^{(B)}.
\end{align}
We then perform a canonical transformation
\begin{align}
\begin{aligned}
    H' &= e^{-\eta S}H\,e^{\eta S} = H + \eta\comm{H}{S} + \frac{\eta^2}{2!}\comm{\comm{H}{S}}{S} + \mathcal{O}(\eta^3)\\
    &= H_0 + \eta\Big(H_1 + \comm{H_0}{S}\Big) + \eta^2\Big(\comm{H_1}{S} + \frac{1}{2}\comm{\comm{H_0}{S}}{S} \Big)\\
    &+ \mathcal{O}(\eta^3),
\end{aligned}
\end{align}
where we choose $\eta S = \eta S^{(A)} + \eta S^{(B)}$ such that we have
\begin{align}
    \eta H_1^{(L)} + \comm{H_0}{\eta S^{(L)}} = 0.
\label{can_constraint}
\end{align}
The result is then 
\begin{align}
    H' = H_0 + \frac{1}{2}\sum_{LL'}\comm{\eta H_1^{(L)}}{\eta S^{(L')}} + \mathcal{O}(\eta^3).
\end{align}
where $L\in \{A,B\}$. We then make appropriate choices for $\eta S^{A}$ and $\eta S^{B}$ \cite{Kittel1963}, compute the commutators, consider that we have condensation of $\uparrow$-bosons, and restrict ourselves to BCS-type scattering processes where the two incoming, as well as outgoing, electrons have opposite momenta. See Appendix \ref{App} for details. The pairing Hamiltonian then takes the form 

\begin{align}
    H_{\rm{pair}} = \sum_{\bm{k}\bm{k}'}V_{\bm{k}\bm{k}'}c^{\dagger}_{\bm{k}\uparrow}c^{\dagger}_{-\bm{k}\downarrow}c_{-\bm{k}'\downarrow}c_{\bm{k}'\uparrow},
    \label{eq:H_pair}
\end{align}

where 

\begin{align}
\begin{aligned}
    V_{\bm{k}\bm{k}'} &= -V^2\Tilde{Q}_0\frac{2\omega_{\bm{k} + \bm{k}'\downarrow}}{(\epsilon_{\bm{k}'} - \epsilon_{\bm{k}})^2 - \omega_{\bm{k} + \bm{k}'\downarrow}^2}A(\bm{k} + \bm{k}',\Omega)\\
    &- \frac{V^2}{N}{\sum_{\bm{h}\in \rm{RBZ}}}\!\!\!' \,B(\bm{k} + \bm{k}'+ \bm{h},\bm{h},\Omega)\\
    &\times\frac{2\big(\omega_{\bm{k} + \bm{k}' + \bm{h}\uparrow} + \omega_{\bm{h}\downarrow}\big)}{\big(\epsilon_{\bm{k}'} - \epsilon_{\bm{k}}\big)^2 - \big(\omega_{\bm{k} + \bm{k}' + \bm{h}\uparrow} + \omega_{\bm{h}\downarrow} \big)^2},
\end{aligned}
\end{align}

and we have defined

\begin{align}
    A(\bm{q},\Omega) &= \frac{1}{2}\big(\Omega^2 + 1\big)\big(u^2_{\bm{q}\downarrow} + v^2_{\bm{q}\downarrow}\big) - 2 \,\Omega \,u_{\bm{q}\downarrow}v_{\bm{q}\downarrow},
\end{align}

as well as 

\begin{align}
\begin{aligned}
    B(\bm{q},\bm{h},\Omega) &= \frac{1}{2}\big(\Omega^2 + 1\big)\big( u^2_{\bm{q}\uparrow}v^2_{\bm{h}\downarrow} +  v^2_{\bm{q}\uparrow}u^2_{\bm{h}\downarrow} \big)\\
    &- 2\,\Omega\, u_{\bm{q}\uparrow}v_{\bm{h}\downarrow}v_{\bm{q}\uparrow}u_{\bm{h}\downarrow}.
\end{aligned}
\end{align}
Here, we have introduced $V \equiv 2\bar{J}/\sqrt{N}$. The two momenta in the sum in Eq.\! \eqref{eq:H_pair} are restricted such that the separation between them is limited to a momentum living in the reduced Brillouin zone of the sublattices. The $A$-factor is the same function as in Ref.\! \cite{Erlandsen2019}, but due to different choices for the sign of the coherence factors, the sign in front of the $u_{\bm{q}\downarrow}v_{\bm{q}\downarrow}$ term is negative instead of positive in this case. Both coherence factors are, in this case, positive for small momenta for both the Néel and stripe phase. For $\Omega = 0$, the $A$-factor then grows large, while for $\Omega = 1$, there is a near-cancellation between the positive and negative contributions to the $A$-factor. The interaction strength is therefore enhanced in the case of asymmetric coupling to the two sublattices, just like in Ref.\! \cite{Erlandsen2019}. The second part of the interaction potential includes a sum over momenta that covers the reduced Brillouin zone of the sublattices, apart from the origin. This term displays similar behavior as the first term in the interaction potential when $\Omega$ is varied. Importantly, the above expressions are valid for both the Néel and stripe phase, meaning that the previously reported enhancement of the critical temperature when coupling asymmetrically to the two sublattices of the AFMI should be expected also for the stripe phase.\\
\indent Examining the first part of the interaction potential, we see that the $\uparrow$-bosons that condense are no longer present. Comparing with the earlier obtained spin-wave result, it is then clear that the magnon energies and coherence factors from Ref.\! \cite{Erlandsen2019} have been replaced by the energies and coherence factors of Schwinger bosons with spin $\sigma = \downarrow$. Moreover, the prefactor $V^2 \tilde{Q}_0$ is the same as the one in Ref.\! \cite{Erlandsen2019}, except that the AFMI spin quantum number $S$ (representing the sublattice magnetization) has been replaced by $\tilde{Q}_0$ which is closely related to the sublattice magnetization $m_A$. As the sublattice magnetization, and therefore $\tilde{Q}_0$, is reduced by frustration, this replacement represents a correction that can influence how the superconductivity depends on the introduction of frustration. In addition, the Schwinger boson energies and coherence factors also depend on $m_A$ (instead of $S$), which will depend on the value of $J_2$ in this treatment of the system.

\section{Gap equation}\label{Section:Gap_equation}

\indent Performing a standard weak-coupling treatment of the superconducting instability \cite{Sigrist1991}, both terms in the interaction potential are found to be attractive for $S_z = 0$ spin-triplet pairing \cite{Erlandsen2019}, with gap function 

\begin{align}
    \Delta_{\bm{k}} = -\sum_{\bm{k}'}V_{\bm{k}\bm{k}',O(\bm{k})}\langle c_{-\bm{k}'\uparrow}c_{\bm{k}'\downarrow} + c_{-\bm{k}'\downarrow}c_{\bm{k}'\uparrow}\rangle/2.
\end{align}
Here $V_{\bm{k}\bm{k}',O(\bm{k})} = \frac{1}{2}(V_{\bm{k}\bm{k}'} - V_{-\bm{k},\bm{k}'})$ is the part of the effective interaction potential that is odd in momentum. The resulting gap equation takes the form \cite{Sigrist1991}
\begin{align}
    \Delta_{\bm{k}} = -\sum_{\bm{k}'}V_{\bm{k}\bm{k}',O(\bm{k})}\frac{\Delta_{\bm{k}'}}{2E_{\bm{k}'}}\tanh(\frac{E_{\bm{k}'}}{2k_B T}),
\end{align}
\noindent where $E_{\bm{k}} = \sqrt{\epsilon^2_k + |\Delta_{\bm{k}}|^2}$, $k_B$ is the Boltzmann constant, and $T$ is the temperature. In order to determine the critical temperature, we consider the linearized gap equation and compute a Fermi surface average 
\begin{align}
    \lambda \Delta_{\bm{k}} = - D_{0} \langle V_{\bm{k}\bm{k}',O(\bm{k})} \Delta_{\bm{k}'} \rangle_{\bm{k}', \text{FS}}.
    \label{eq:lambda}
\end{align}
The critical temperature is then given by \cite{Sigrist1991}
\begin{align}
    k_B T_c = 1.14\,\omega_c\,e^{-1/\lambda},
\end{align}
where $D_0$ is the density of states at the the Fermi level, $\omega_c$ is the boson spectrum cutoff, and the dimensionless coupling constant $\lambda$ is the largest eigenvalue of the eigenvalue problem in Eq.\! \eqref{eq:lambda}.\\
\indent The eigenvalue problem can be treated numerically by picking discrete points on the Fermi surface and solving the resulting matrix eigenvalue problem using a linear algebra library \cite{Armadillo1, Armadillo2}. The density of states at the Fermi level is obtained from numerical evaluation of the elliptical integral derived in Ref.\! \cite{Piasecki2008}. The second part of the interaction potential, involving the $B$-factor, is found to have little influence on the dimensionless coupling constant and the critical temperature. Calculating this part of the potential is computationally costly as an integral over the Brillouin zone then needs to be computed for each independent set of momenta $\bm{k}$, $\bm{k}'$ in Eq.\! \eqref{eq:lambda}. In order to increase the momentum-space resolution, the following results are therefore obtained without the second part of the interaction potential. 

\section{Results}\label{Section:Results}

\begin{figure}[ht]  
    \begin{center}
        \hspace{-1.1cm}
        \includegraphics[width=0.65\columnwidth,trim= 0.1cm 0.5cm 0.8cm 0.5cm,clip=true]{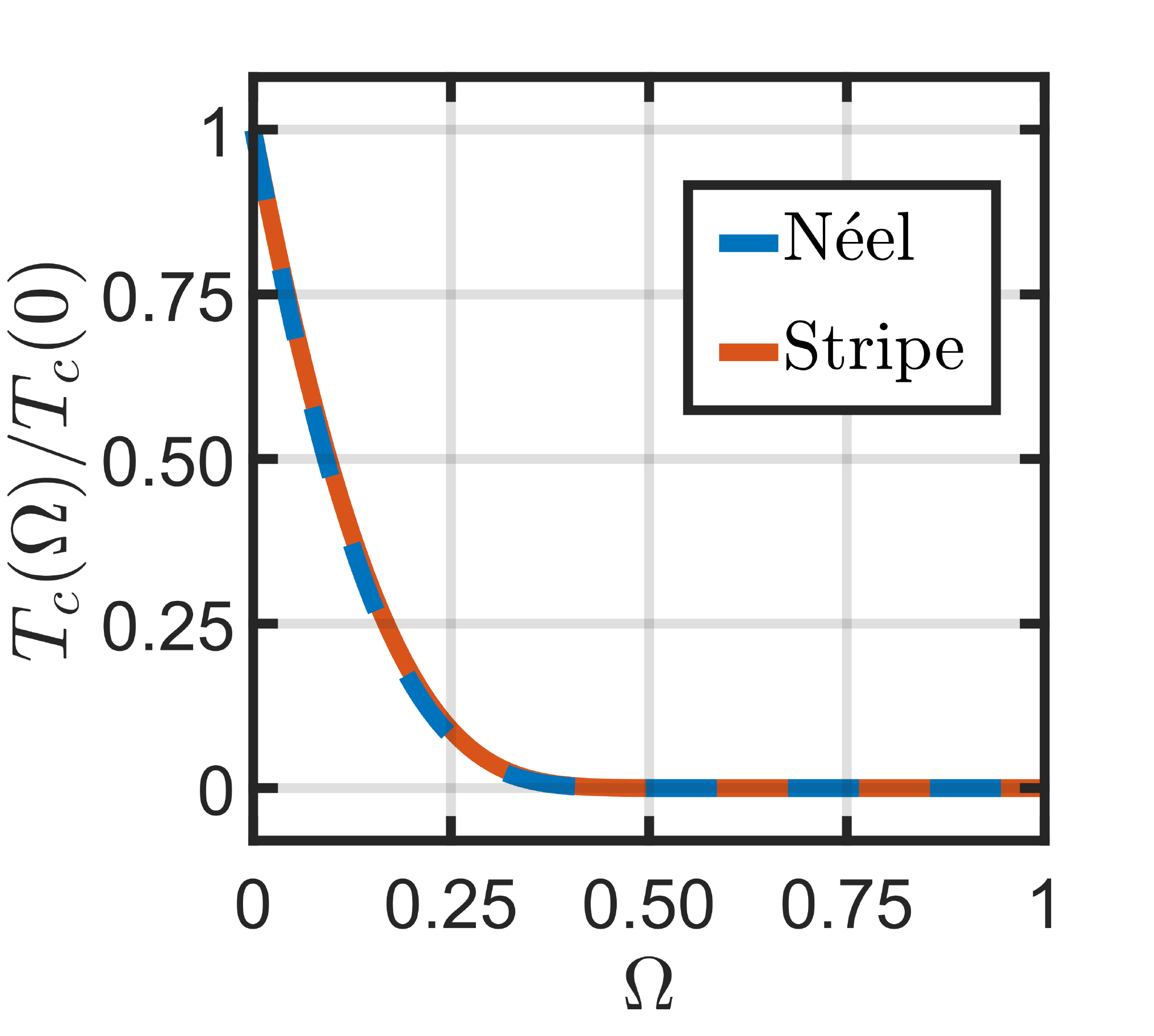}
    \end{center}
    \caption{Superconducting critical temperature $T_c$ presented as a function of the coupling asymmetry parameter $\Omega$ for $J_1 = 5 \,\rm{meV}$, $K = J_1/4000$, $S=1$, $t = 0.8 \,\rm{eV}$, $\mu = -3.5 t$, and $\bar{J} = 13 \,\rm{meV}$. The critical temperature has been normalized by its value for $\Omega = 0$ for the Néel phase ($J_2 = 0.5J_1$) and the stripe phase ($J_2 = 0.6 J_1$), respectively.}
    \label{fig:Tc_Omega}
\end{figure}

The critical temperature as a function of the asymmetry parameter $\Omega$ is presented in Fig.\! \ref{fig:Tc_Omega} for both the Néel and stripe phase. As expected, based on the discussion in Sec.\! \ref{Section:Eff_int}, the critical temperature rises up when $\Omega \rightarrow 0$ (uncompensated interface), as the $A$-factor grows larger. The displayed values for each magnetic phase have been normalized by the critical temperature corresponding to the same magnetic phase and $\Omega = 0$. Based on the result that an uncompensated interface provides an enhancement of the critical temperature for both magnetic phases, we focus on the case of $\Omega = 0$ in the following. 

\begin{figure}[ht] 
    \begin{center}
    \hspace{-0.5cm}
        \includegraphics[width=0.7\columnwidth,trim= 1.0cm 1.5cm 3.3cm 0.5cm,clip=true]{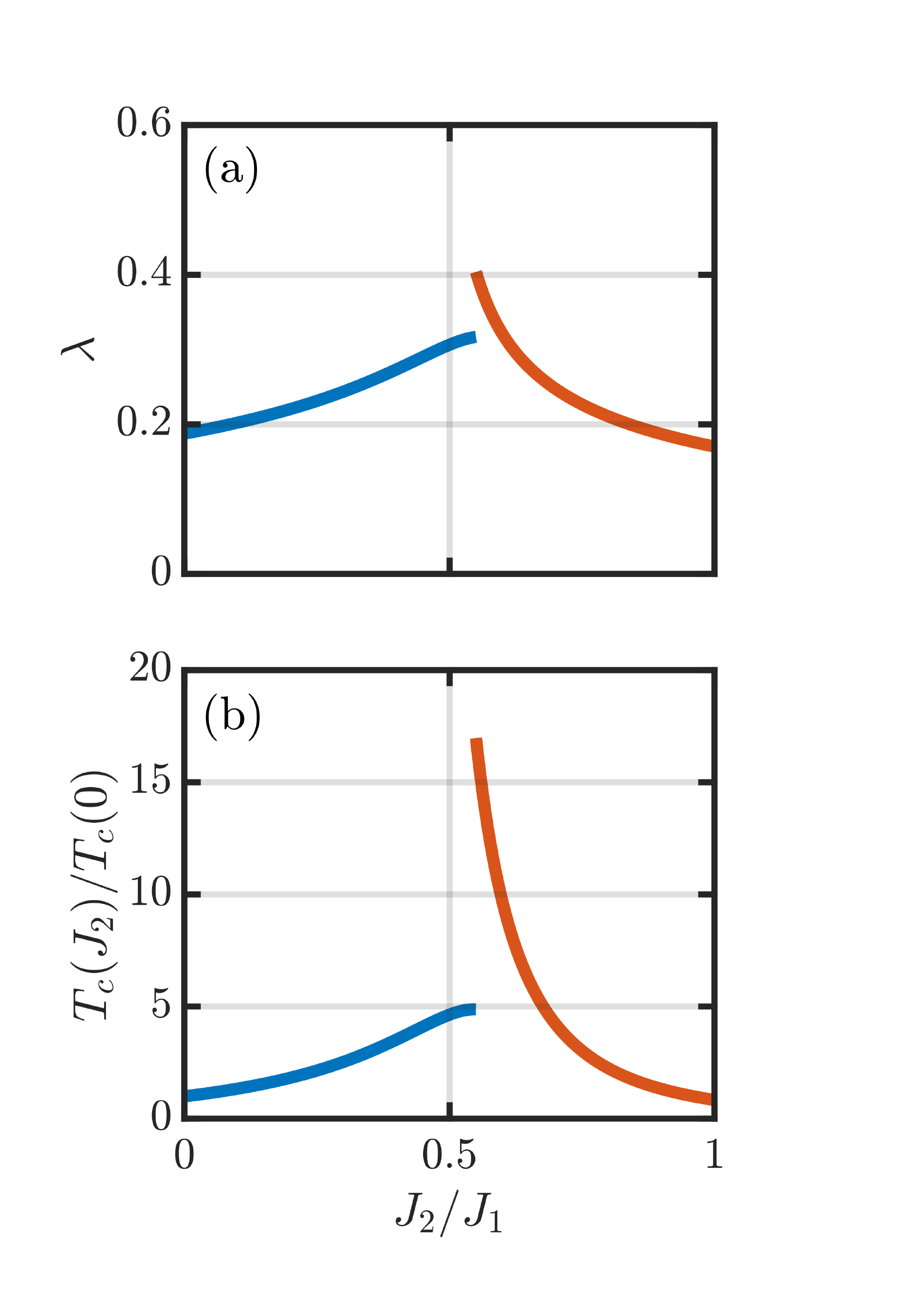}
    \end{center}
    \caption{Dimensionless coupling constant $\lambda$ (a) and superconducting critical temperature $T_c$ (b) presented as a function of the next-nearest neighbor exchange coupling $J_2$ in the antiferromagnet for $J_1 = 5 \,\rm{meV}$, $K = J_1/4000$, $S=1$, $t = 0.8 \,\rm{eV}$, $\mu = -3.5 t$, $\bar{J} = 13 \,\rm{meV}$, and $\Omega = 0$.}
    \label{fig:Tc_J2}
\end{figure} 

Next, we investigate how the dimensionless coupling constant and the critical temperature depend on the next-nearest neighbor interaction $J_2$ in the AFMI. These results are displayed in Fig.\! \ref{fig:Tc_J2}. For both magnetic phases, we find that approaching the phase transition leads to a larger dimensionless coupling constant and critical temperature. For the Néel phase, the introduction of frustration gives rise to a smaller cutoff on the boson spectrum (Fig.\! \ref{fig:Neel_disp_rel}), and a smaller sublattice magnetization which leads to a reduction of the prefactor $\tilde{Q}_0$ (Fig.\! \ref{fig:MF_params}) in the interaction potential. On the other hand, the frustration produces larger coherence factors (Fig.\! \ref{fig:Neel_u}) and a flatter boson dispersion relation, which enters in the denominator of the interaction potential. The overall effect is that the dimensionless coupling constant increases, which leads to a rise in the critical temperature despite the reduction in the boson spectrum cutoff. In the vicinity of the transition to the stripe phase, the $T_c$ curve becomes flatter as the factor $\tilde{Q}_0$ drops more quickly. For smaller(larger) AFMI spin quantum number $S$, $\tilde{Q}_0$ will be reduced more(less) dramatically as one approaches the transition point. For the spin $S=1/2$ case, where the long-range order can vanish for sufficiently strong frustration, the critical temperature resulting from the above calculation will dive down. However, for a three-dimensional system, with stronger tendency of ordering, the magnetization will generally be reduced less than for the two-dimensional model system considered here. The typical result for an actual three-dimensional AFMI with an uncompensated interface (potentially excluding the case of spin-$1/2$ on a simple cubic lattice \cite{Farnell2016}), is then expected to be similar to the above result (Fig.\! \ref{fig:Tc_J2}) where the dimensionless coupling constant increases as one approaches the stripe phase, leading to a higher critical temperature.\\
\indent For the stripe phase, approaching the transition to the Néel phase, the cutoff on the boson spectrum (Fig.\! \ref{fig:Stripe_disp_rel}) and the factor $\tilde{Q}_0$ (Fig.\! \ref{fig:MF_params}) are once again reduced. In addition, the maximum value of the coherence factors also decreases (Fig.\! \ref{fig:Stripe_u}), in contrast to the Néel case. This could indicate that the induced electron-electron interactions are becoming weaker, potentially leading to a smaller $\lambda$ and $T_c$. It is, however, the case that the region in $k$-space where the coherence factors take on large values is stretched out in the direction of the stripes (the $k_y$-direction), followed by a flattening of the dispersion relation in this direction. In order to take advantage of favorable scattering processes in the $k_y$-direction, while keeping the involved electrons on the Fermi surface, the magnitude of the gap function is shifted towards the $k_x$-axis compared to the standard $p$-wave gap function of the unfrustrated Néel state \cite{Erlandsen2019}. More scattering processes with large and moderately large contributions compensate for the reduction, instead of increase, in the maximum value of the coherence factors. The dimensionless coupling constant therefore still grows as one approaches the phase transition. As the sublattice magnetization is more robust for the stripe phase than the Néel phase (Fig.\! \ref{fig:MF_params}), the maximum value of $\lambda$ actually ends up being slightly higher for the stripe phase due to a larger value for $\tilde{Q}_0$. As the critical temperature is very sensitive to $\lambda$, the critical temperature rises up quite dramatically on the stripe side of the transition in our calculation. The main conclusion from the stripe phase is that increasing the fluctuations by approaching the transition to the Néel phase can be favorable for the superconductivity and that the more stable sublattice magnetization of the stripe phase can be an advantage.

\section{Summary}\label{Section:Summary}
We have investigated superconductivity in a normal metal, induced by an interfacial coupling to an antiferromagnetic insulator. We have shown that next-nearest neighbor frustration in a Néel antiferromagnet with an uncompensated interface can lead to an enhancement of the superconducting critical temperature. Moreover, coupling to an uncompensated antiferromagnetic interface is found to be favorable for the superconductivity regardless of whether the antiferromagnet is in a Néel phase or a stripe phase. For the stripe phase, arising from large next-nearest neighbor interaction in the antiferromagnet, we find that amplifying the magnetic fluctuations by approaching the transition to the Néel phase, once again, can lead to a rise in the critical temperature.      

\section{Acknowledgements}
We thank Even Thingstad for valuable discussions. We acknowledge financial support from the Research Council of Norway Grant No.\! 262633 “Center of Excellence on Quantum Spintronics” and Grant No.\! 250985, “Fundamentals of Low-dissipative Topological Matter”.

\appendix
\section{Derivation of the effective interaction} \label{App}

In order to obtain an effective theory of interacting electrons, we choose 

\begin{align}
\begin{aligned}
    \eta S^{(A)} &= -\frac{2\Bar{J}\,\Omega}{N}\,\sum_{\bm{q}\in \Box}\sum_{\substack{\bm{k}\bm{k}'\\ \in \text{RBZ}}}\Big[\big(x^{\alpha,\uparrow\downarrow}_{\bm{k},\bm{k}'\bm{q}}u_{\bm{k}\uparrow} u_{\bm{k}'\downarrow}\alpha^{\dagger}_{\bm{k}\uparrow} \alpha_{\bm{k}'\downarrow}\\
    &+ y^{\beta,\downarrow\uparrow}_{\bm{k},\bm{k}'\bm{q}}v_{\bm{k}\uparrow}v_{\bm{k}'\downarrow}  \beta^{\dagger}_{-\bm{k}'\downarrow}\beta_{-\bm{k}\uparrow} - z^{-,\uparrow\downarrow}_{\bm{k},\bm{k}'\bm{q}}u_{\bm{k}\uparrow}v_{\bm{k}'\downarrow} \alpha^{\dagger}_{\bm{k}\uparrow} \beta^{\dagger}_{-\bm{k}'\downarrow}\\
    &- w^{+,\downarrow\uparrow}_{\bm{k},\bm{k}'\bm{q}}v_{\bm{k}\uparrow}u_{\bm{k}'\downarrow} \alpha_{\bm{k}'\downarrow}\beta_{-\bm{k}\uparrow}\big)c^{\dagger}_{\bm{q}\downarrow}c_{\bm{k} + \bm{q} - \bm{k}'\uparrow}\\
    & + \big(x^{\alpha,\downarrow\uparrow}_{\bm{k},\bm{k}'\bm{q}}u_{\bm{k}\downarrow}u_{\bm{k}'\uparrow} \alpha^{\dagger}_{\bm{k}\downarrow} \alpha_{\bm{k}'\uparrow} + y^{\beta,\uparrow\downarrow}_{\bm{k},\bm{k}'\bm{q}}v_{\bm{k}\downarrow}v_{\bm{k}'\uparrow}  \beta^{\dagger}_{-\bm{k}'\uparrow}\beta_{-\bm{k}\downarrow}\\
    &- z^{-,\downarrow\uparrow}_{\bm{k},\bm{k}'\bm{q}}u_{\bm{k}\downarrow}v_{\bm{k}'\uparrow}  \alpha^{\dagger}_{\bm{k}\downarrow}\beta^{\dagger}_{-\bm{k}'\uparrow}\\
    &- w^{+,\uparrow\downarrow}_{\bm{k},\bm{k}'\bm{q}} v_{\bm{k}\downarrow}u_{\bm{k}'\uparrow} \alpha_{\bm{k}'\uparrow}\beta_{-\bm{k}\downarrow} \big)c^{\dagger}_{\bm{q}\uparrow}c_{\bm{k} + \bm{q} - \bm{k}'\downarrow}\Big],
\end{aligned}
\end{align}

and 

\begin{align}
\begin{aligned}
    \eta S^{(B)} &= \frac{2\Bar{J}}{N}\,\sum_{\bm{q}\in \Box}\sum_{\substack{\bm{k}\bm{k}'\\ \in \rm{RBZ}}}\Big[\big(y^{\alpha,\uparrow\downarrow}_{\bm{k},\bm{k}'\bm{q}}v_{\bm{k}\downarrow}v_{\bm{k}'\uparrow}  \alpha^{\dagger}_{-\bm{k}'\uparrow}\alpha_{-\bm{k}\downarrow}\\
    &+ x^{\beta,\downarrow\uparrow}_{\bm{k},\bm{k}'\bm{q}}u_{\bm{k}\downarrow}u_{\bm{k}'\uparrow}\beta^{\dagger}_{\bm{k}\downarrow}\beta_{\bm{k}'\uparrow} - z^{+,\downarrow\uparrow}_{\bm{k},\bm{k}'\bm{q}}v_{\bm{k}\downarrow}u_{\bm{k}'\uparrow} \alpha_{-\bm{k}\downarrow}\beta_{\bm{k}'\uparrow}\\
    &- w^{-,\uparrow\downarrow}_{\bm{k},\bm{k}'\bm{q}}u_{\bm{k}\downarrow}v_{\bm{k}'\uparrow}\alpha^{\dagger}_{-\bm{k}'\uparrow}\beta^{\dagger}_{\bm{k}\downarrow} \big)c^{\dagger}_{\bm{q}\downarrow}c_{\bm{k} + \bm{q} - \bm{k}'\uparrow}\\
    &+ \big(y^{\alpha,\downarrow\uparrow}_{\bm{k},\bm{k}'\bm{q}}v_{\bm{k}\uparrow} v_{\bm{k}'\downarrow} \alpha^{\dagger}_{-\bm{k}'\downarrow}\alpha_{-\bm{k}\uparrow} + x^{\beta,\uparrow\downarrow}_{\bm{k},\bm{k}'\bm{q}}u_{\bm{k}\uparrow}u_{\bm{k}'\downarrow}\beta^{\dagger}_{\bm{k}\uparrow}\beta_{\bm{k}'\downarrow}\\
    &- z^{+,\uparrow\downarrow}_{\bm{k},\bm{k}'\bm{q}}v_{\bm{k}\uparrow}u_{\bm{k}'\downarrow} \alpha_{-\bm{k}\uparrow}\beta_{\bm{k}'\downarrow}\\
    &- w^{-,\downarrow\uparrow}_{\bm{k},\bm{k}'\bm{q}}u_{\bm{k}\uparrow}v_{\bm{k}'\downarrow} \alpha^{\dagger}_{-\bm{k}'\downarrow}\beta^{\dagger}_{\bm{k}\uparrow}\big)c^{\dagger}_{\bm{q}\uparrow}c_{\bm{k} + \bm{q} - \bm{k}'\downarrow}\Big],
\end{aligned}
\end{align}

where 

\begin{subequations}
\begin{align}
    &x^{\alpha,\uparrow\downarrow}_{\bm{k},\bm{k}'\bm{q}} = \frac{1}{\epsilon_{\bm{k} + \bm{q} - \bm{k}'} - \epsilon_{\bm{q}} + \omega_{\bm{k}'\downarrow\alpha} - \omega_{\bm{k}\uparrow\alpha}},\\
    &y^{\beta,\downarrow\uparrow}_{\bm{k},\bm{k}'\bm{q}} = \frac{1}{\epsilon_{\bm{k} + \bm{q} - \bm{k}'} - \epsilon_{\bm{q}} + \omega_{\bm{k}\uparrow\beta} - \omega_{\bm{k}'\downarrow\beta}},\\
    &z^{-,\uparrow\downarrow}_{\bm{k},\bm{k}'\bm{q}} = \frac{1}{\epsilon_{\bm{k} + \bm{q} - \bm{k}'} - \epsilon_{\bm{q}} - \omega_{\bm{k}\uparrow\alpha} - \omega_{\bm{k}'\downarrow\beta}},\\
    &w^{+,\downarrow\uparrow}_{\bm{k},\bm{k}'\bm{q}} = \frac{1}{\epsilon_{\bm{k} + \bm{q} - \bm{k}'} - \epsilon_{\bm{q}} + \omega_{\bm{k}'\downarrow\alpha} + \omega_{\bm{k}\uparrow\beta}},
\end{align}
\end{subequations}

and e.g.\! 

\begin{subequations}
\begin{align}
    &z^{+,\uparrow\downarrow}_{\bm{k},\bm{k}'\bm{q}} = \frac{1}{\epsilon_{\bm{k} + \bm{q} - \bm{k}'} - \epsilon_{\bm{q}} + \omega_{\bm{k}\uparrow\alpha} + \omega_{\bm{k}'\downarrow\beta}},\\
    &w^{-,\downarrow\uparrow}_{\bm{k},\bm{k}'\bm{q}} = \frac{1}{\epsilon_{\bm{k} + \bm{q} - \bm{k}'} - \epsilon_{\bm{q}} - \omega_{\bm{k}'\downarrow\alpha} - \omega_{\bm{k}\uparrow\beta}}.
\end{align}
\end{subequations}
Prior to commencing a calculation of the commutators, it turns out to be advantageous to further split up the terms in the interaction Hamiltonian 

\begin{align}
    H^{(L)}_{\text{int}} = H^{(L,+)}_{\text{int}} + H^{(L,-)}_{\text{int}},
\end{align}

where 

\begin{align}
\begin{aligned}
    H^{(A,+)}_{\text{int}} &= -\frac{2\Bar{J}\,\Omega}{N}\,\sum_{\bm{q}\in \Box}\sum_{\substack{\bm{k}\bm{k}'\\ \in \rm{RBZ}}}\big(u_{\bm{k}\uparrow} u_{\bm{k}'\downarrow}\alpha^{\dagger}_{\bm{k}\uparrow} \alpha_{\bm{k}'\downarrow}\\
    &+ v_{\bm{k}\uparrow}v_{\bm{k}'\downarrow} \beta_{-\bm{k}\uparrow} \beta^{\dagger}_{-\bm{k}'\downarrow} - u_{\bm{k}\uparrow}v_{\bm{k}'\downarrow} \alpha^{\dagger}_{\bm{k}\uparrow} \beta^{\dagger}_{-\bm{k}'\downarrow}\\
    &- v_{\bm{k}\uparrow}u_{\bm{k}'\downarrow} \beta_{-\bm{k}\uparrow} \alpha_{\bm{k}'\downarrow}\big)c^{\dagger}_{\bm{q}\downarrow}c_{\bm{k} + \bm{q} - \bm{k}'\uparrow},
\end{aligned}
\end{align}

\begin{align}
\begin{aligned}
    H^{(A,-)}_{\text{int}} &= -\frac{2\Bar{J}\,\Omega}{N}\,\sum_{\bm{q}\in \Box}\sum_{\substack{\bm{k}\bm{k}'\\ \in \rm{RBZ}}}\big(u_{\bm{k}\downarrow}u_{\bm{k}'\uparrow} \alpha^{\dagger}_{\bm{k}\downarrow} \alpha_{\bm{k}'\uparrow}\\
    &+ v_{\bm{k}\downarrow}v_{\bm{k}'\uparrow} \beta_{-\bm{k}\downarrow} \beta^{\dagger}_{-\bm{k}'\uparrow} - u_{\bm{k}\downarrow}v_{\bm{k}'\uparrow}  \alpha^{\dagger}_{\bm{k}\downarrow}\beta^{\dagger}_{-\bm{k}'\uparrow}\\
    &- v_{\bm{k}\downarrow}u_{\bm{k}'\uparrow} \beta_{-\bm{k}\downarrow} \alpha_{\bm{k}'\uparrow}\big)c^{\dagger}_{\bm{q}\uparrow}c_{\bm{k} + \bm{q} - \bm{k}'\downarrow},
\end{aligned}
\end{align}

\begin{align}
\begin{aligned}
    H^{(B,+)}_{\text{int}} &= \frac{2\Bar{J}}{N}\,\sum_{\bm{q}\in \Box}\sum_{\substack{\bm{k}\bm{k}'\\ \in \rm{RBZ}}}\big(v_{\bm{k}\downarrow}v_{\bm{k}'\uparrow} \alpha_{-\bm{k}\downarrow} \alpha^{\dagger}_{-\bm{k}'\uparrow}\\
    &+ u_{\bm{k}\downarrow}u_{\bm{k}'\uparrow}\beta^{\dagger}_{\bm{k}\downarrow}\beta_{\bm{k}'\uparrow} - v_{\bm{k}\downarrow}u_{\bm{k}'\uparrow} \alpha_{-\bm{k}\downarrow}\beta_{\bm{k}'\uparrow}\\
    &- u_{\bm{k}\downarrow}v_{\bm{k}'\uparrow}\beta^{\dagger}_{\bm{k}\downarrow} \alpha^{\dagger}_{-\bm{k}'\uparrow}\big)c^{\dagger}_{\bm{q}\downarrow}c_{\bm{k} + \bm{q} - \bm{k}'\uparrow},
\end{aligned}
\end{align}

\begin{align}
\begin{aligned}
    H^{(B,-)}_{\text{int}} &= \frac{2\Bar{J}}{N}\,\sum_{\bm{q}\in \Box}\sum_{\substack{\bm{k}\bm{k}'\\ \in \rm{RBZ}}}\big(v_{\bm{k}\uparrow} v_{\bm{k}'\downarrow}\alpha_{-\bm{k}\uparrow} \alpha^{\dagger}_{-\bm{k}'\downarrow}\\
    &+ u_{\bm{k}\uparrow}u_{\bm{k}'\downarrow}\beta^{\dagger}_{\bm{k}\uparrow}\beta_{\bm{k}'\downarrow} - v_{\bm{k}\uparrow}u_{\bm{k}'\downarrow} \alpha_{-\bm{k}\uparrow}\beta_{\bm{k}'\downarrow}\\
    &- u_{\bm{k}\uparrow}v_{\bm{k}'\downarrow}\beta^{\dagger}_{\bm{k}\uparrow} \alpha^{\dagger}_{-\bm{k}'\downarrow}\big)c^{\dagger}_{\bm{q}\uparrow}c_{\bm{k} + \bm{q} - \bm{k}'\downarrow}.
\end{aligned}
\end{align}
When calculating the boson commutators, leaving us with four electron operators and two boson operators, and exchanging the pair of remaining boson operators with its ground state expectation value, we find that only processes where the two incoming electrons have opposite spins give non-zero contributions. These processes then conserve the electron spin and are of the same type as the processes one obtains from a Holstein-Primakoff treatment of the magnetic system \cite{Fjaerbu2018, Fjaerbu2019, Erlandsen2019}. The pairing Hamiltonian can then be written as  

\begin{align}
    H_{\rm{pair}} = \frac{1}{2}\sum_{LL'\sigma}\comm{H_{\rm{int}}^{(L,\sigma)}}{\eta S^{(L',-\sigma)}},
\end{align}
where 

\begin{align}
\begin{aligned}
    \eta S^{(A,+)} &= -\frac{2\Bar{J}\,\Omega}{N}\,\sum_{\bm{q}\in \Box}\sum_{\substack{\bm{k}\bm{k}'\\ \in \rm{RBZ}}}\big(x^{\alpha,\uparrow\downarrow}_{\bm{k},\bm{k}'\bm{q}}u_{\bm{k}\uparrow} u_{\bm{k}'\downarrow}\alpha^{\dagger}_{\bm{k}\uparrow} \alpha_{\bm{k}'\downarrow}\\
    &+ y^{\beta,\downarrow\uparrow}_{\bm{k},\bm{k}'\bm{q}}v_{\bm{k}\uparrow}v_{\bm{k}'\downarrow}  \beta^{\dagger}_{-\bm{k}'\downarrow}\beta_{-\bm{k}\uparrow} - z^{-,\uparrow\downarrow}_{\bm{k},\bm{k}'\bm{q}}u_{\bm{k}\uparrow}v_{\bm{k}'\downarrow} \alpha^{\dagger}_{\bm{k}\uparrow} \beta^{\dagger}_{-\bm{k}'\downarrow}\\
    &- w^{+,\downarrow\uparrow}_{\bm{k},\bm{k}'\bm{q}}v_{\bm{k}\uparrow}u_{\bm{k}'\downarrow} \alpha_{\bm{k}'\downarrow}\beta_{-\bm{k}\uparrow}\big)c^{\dagger}_{\bm{q}\downarrow}c_{\bm{k} + \bm{q} - \bm{k}'\uparrow},
\end{aligned}
\end{align}

\begin{align}
\begin{aligned}
    \eta S^{(A,-)} &= -\frac{2\Bar{J}\,\Omega}{N}\,\sum_{\bm{q}\in \Box}\sum_{\substack{\bm{k}\bm{k}'\\ \in \rm{RBZ}}}\big(x^{\alpha,\downarrow\uparrow}_{\bm{k},\bm{k}'\bm{q}}u_{\bm{k}\downarrow}u_{\bm{k}'\uparrow} \alpha^{\dagger}_{\bm{k}\downarrow} \alpha_{\bm{k}'\uparrow}\\
    &+ y^{\beta,\uparrow\downarrow}_{\bm{k},\bm{k}'\bm{q}}v_{\bm{k}\downarrow}v_{\bm{k}'\uparrow}  \beta^{\dagger}_{-\bm{k}'\uparrow}\beta_{-\bm{k}\downarrow}- z^{-,\downarrow\uparrow}_{\bm{k},\bm{k}'\bm{q}}u_{\bm{k}\downarrow}v_{\bm{k}'\uparrow}  \alpha^{\dagger}_{\bm{k}\downarrow}\beta^{\dagger}_{-\bm{k}'\uparrow}\\
    & - w^{+,\uparrow\downarrow}_{\bm{k},\bm{k}'\bm{q}} v_{\bm{k}\downarrow}u_{\bm{k}'\uparrow} \alpha_{\bm{k}'\uparrow}\beta_{-\bm{k}\downarrow} \big)c^{\dagger}_{\bm{q}\uparrow}c_{\bm{k} + \bm{q} - \bm{k}'\downarrow},
\end{aligned}
\end{align}

\begin{align}
\begin{aligned}
    \eta S^{(B,+)} &= \frac{2\Bar{J}}{N}\,\sum_{\bm{q}\in \Box}\sum_{\substack{\bm{k}\bm{k}'\\ \in \rm{RBZ}}}\big(y^{\alpha,\uparrow\downarrow}_{\bm{k},\bm{k}'\bm{q}}v_{\bm{k}\downarrow}v_{\bm{k}'\uparrow}  \alpha^{\dagger}_{-\bm{k}'\uparrow}\alpha_{-\bm{k}\downarrow}\\
    &+ x^{\beta,\downarrow\uparrow}_{\bm{k},\bm{k}'\bm{q}}u_{\bm{k}\downarrow}u_{\bm{k}'\uparrow}\beta^{\dagger}_{\bm{k}\downarrow}\beta_{\bm{k}'\uparrow}- z^{+,\downarrow\uparrow}_{\bm{k},\bm{k}'\bm{q}}v_{\bm{k}\downarrow}u_{\bm{k}'\uparrow} \alpha_{-\bm{k}\downarrow}\beta_{\bm{k}'\uparrow}\\
    & - w^{-,\uparrow\downarrow}_{\bm{k},\bm{k}'\bm{q}}u_{\bm{k}\downarrow}v_{\bm{k}'\uparrow}\alpha^{\dagger}_{-\bm{k}'\uparrow}\beta^{\dagger}_{\bm{k}\downarrow} \big)c^{\dagger}_{\bm{q}\downarrow}c_{\bm{k} + \bm{q} - \bm{k}'\uparrow},
\end{aligned}
\end{align}

\begin{align}
\begin{aligned}
    \eta S^{(B,-)} &= \frac{2\Bar{J}}{N}\,\sum_{\bm{q}\in \Box}\sum_{\substack{\bm{k}\bm{k}'\\ \in \rm{RBZ}}}\big(y^{\alpha,\downarrow\uparrow}_{\bm{k},\bm{k}'\bm{q}}v_{\bm{k}\uparrow} v_{\bm{k}'\downarrow} \alpha^{\dagger}_{-\bm{k}'\downarrow}\alpha_{-\bm{k}\uparrow}\\
    &+ x^{\beta,\uparrow\downarrow}_{\bm{k},\bm{k}'\bm{q}}u_{\bm{k}\uparrow}u_{\bm{k}'\downarrow}\beta^{\dagger}_{\bm{k}\uparrow}\beta_{\bm{k}'\downarrow}- z^{+,\uparrow\downarrow}_{\bm{k},\bm{k}'\bm{q}}v_{\bm{k}\uparrow}u_{\bm{k}'\downarrow} \alpha_{-\bm{k}\uparrow}\beta_{\bm{k}'\downarrow}\\
    & - w^{-,\downarrow\uparrow}_{\bm{k},\bm{k}'\bm{q}}u_{\bm{k}\uparrow}v_{\bm{k}'\downarrow} \alpha^{\dagger}_{-\bm{k}'\downarrow}\beta^{\dagger}_{\bm{k}\uparrow}\big)c^{\dagger}_{\bm{q}\uparrow}c_{\bm{k} + \bm{q} - \bm{k}'\downarrow}.
\end{aligned}
\end{align}

Computing the commutators, grouping together terms, and exchanging the boson operator pairs by their ground state expectation value, we obtain 

\begin{align}
    H_{\rm{pair}} = H^{AA} + H^{BB} + H^{AB} + H^{BA},
\end{align}

where 

\begin{align}
\begin{aligned}
    &H^{AA} = \Omega^2 \frac{V^2}{2N}\sum_{\substack{\bm{q}\bm{l}\\\in \Box}}\sum_{\substack{\bm{k}\bm{k}'\\ \in \rm{RBZ}}}\Bigg\{
    \Big(x^{\alpha,\downarrow\uparrow}_{\bm{k}',\bm{k},\bm{l}} - x^{\alpha,\uparrow\downarrow}_{\bm{k},\bm{k}',\bm{q}}\Big)u^2_{\bm{k}\uparrow} u^2_{\bm{k}'\downarrow}\\
    &\times \Big(\langle\alpha^{\dagger}_{\bm{k}\uparrow}\alpha_{\bm{k}\uparrow}\rangle - \langle \alpha^{\dagger}_{\bm{k}'\downarrow}\alpha_{\bm{k}'\downarrow}\rangle\Big) + \Big(y^{\beta,\downarrow\uparrow}_{\bm{k},\bm{k}',\bm{q}} - y^{\beta,\uparrow\downarrow}_{\bm{k}',\bm{k},\bm{l}} \Big)v^2_{\bm{k}\uparrow}v^2_{\bm{k}'\downarrow}\\
    &\times\Big(\langle \beta^{\dagger}_{-\bm{k}\uparrow}\beta_{-\bm{k}\uparrow}\rangle - \langle \beta^{\dagger}_{-\bm{k}'\downarrow}\beta_{-\bm{k}'\downarrow}\rangle\Big) + \Big(z^{-,\uparrow\downarrow}_{\bm{k},\bm{k}',\bm{q}} - w^{+,\uparrow\downarrow}_{\bm{k}',\bm{k},\bm{l}}\Big)\\
    &\times u^2_{\bm{k}\uparrow}v^2_{\bm{k}'\downarrow}\Big(\langle\alpha^{\dagger}_{\bm{k}\uparrow}\alpha_{\bm{k}\uparrow}\rangle+ \langle\beta^{\dagger}_{-\bm{k}'\downarrow}\beta_{-\bm{k}'\downarrow}\rangle + 1\Big) \\
    &+ \Big(z^{-,\downarrow\uparrow}_{\bm{k}',\bm{k},\bm{l}} - w^{+,\downarrow\uparrow}_{\bm{k},\bm{k}',\bm{q}} \Big)v^2_{\bm{k}\uparrow}u^2_{\bm{k}'\downarrow}\Big(\langle\beta^{\dagger}_{-\bm{k}\uparrow}\beta_{-\bm{k}\uparrow}\rangle\\
    & + \langle\alpha^{\dagger}_{\bm{k}'\downarrow}\alpha_{\bm{k}'\downarrow}\rangle+ 1\Big)\Bigg\}c^{\dagger}_{\bm{q}\downarrow}c_{\bm{k} + \bm{q} - \bm{k}'\uparrow}c^{\dagger}_{\bm{l}\uparrow}c_{\bm{k}' + \bm{l} - \bm{k}\downarrow},
\end{aligned}
\end{align}    

\begin{align}
\begin{aligned}
    &H^{BB} = \frac{V^2}{2N}\sum_{\substack{\bm{q}\bm{l}\\\in \Box}}\sum_{\substack{\bm{k}\bm{k}'\\ \in \rm{RBZ}}}\Bigg\{\Big(y^{\alpha,\downarrow\uparrow}_{\bm{k}',\bm{k},\bm{l}} - y^{\alpha,\uparrow\downarrow}_{\bm{k},\bm{k}',\bm{q}}\Big)v^2_{\bm{k}\downarrow}v^2_{\bm{k}'\uparrow}\\
    &\times\Big(\langle\alpha^{\dagger}_{-\bm{k}'\uparrow}\alpha_{-\bm{k}'\uparrow}\rangle - \langle\alpha^{\dagger}_{-\bm{k}\downarrow}\alpha_{-\bm{k}\downarrow}\rangle\Big) +\Big(x^{\beta,\downarrow\uparrow}_{\bm{k},\bm{k}',\bm{q}} - x^{\beta,\uparrow\downarrow}_{\bm{k}',\bm{k},\bm{l}}\Big)\\
    &\times u^2_{\bm{k}\downarrow}u^2_{\bm{k}'\uparrow}\Big(\langle\beta^{\dagger}_{\bm{k}'\uparrow}\beta_{\bm{k}'\uparrow}\rangle - \langle\beta^{\dagger}_{\bm{k}\downarrow}\beta_{\bm{k}\downarrow}\rangle\Big) + \Big(w^{-,\downarrow\uparrow}_{\bm{k}',\bm{k},\bm{l}} - z^{+,\downarrow\uparrow}_{\bm{k},\bm{k}',\bm{q}} \Big)\\
    &\times v^2_{\bm{k}\downarrow}u^2_{\bm{k}'\uparrow}\Big(\langle\beta^{\dagger}_{\bm{k}'\uparrow}\beta_{\bm{k}'\uparrow}\rangle + \langle\alpha^{\dagger}_{-\bm{k}\downarrow}\alpha_{-\bm{k}\downarrow}\rangle + 1\Big)\\
    &+ \Big(w^{-,\uparrow\downarrow}_{\bm{k},\bm{k}',\bm{q}} - z^{+,\uparrow\downarrow}_{\bm{k}',\bm{k},\bm{l}}\Big)u^2_{\bm{k}\downarrow}v^2_{\bm{k}'\uparrow}\Big(\langle\alpha^{\dagger}_{-\bm{k}'\uparrow}\alpha_{-\bm{k}'\uparrow}\rangle\\
    &+ \langle\beta^{\dagger}_{\bm{k}\downarrow}\beta_{\bm{k}\downarrow}\rangle +  1\Big) \Bigg\}c^{\dagger}_{\bm{q}\downarrow}c_{\bm{k} + \bm{q} - \bm{k}'\uparrow}c^{\dagger}_{\bm{l}\uparrow}c_{\bm{k}' + \bm{l} - \bm{k}\downarrow},
\end{aligned}
\end{align}

\begin{align}
\begin{aligned}
    &H^{AB} = -\Omega\frac{V^2}{2N}\sum_{\substack{\bm{q}\bm{l}\\\in \Box}}\sum_{\substack{\bm{k}\bm{k}'\\ \in \rm{RBZ}}}\Bigg\{
     \Big(y^{\alpha,\downarrow\uparrow}_{-\bm{k},-\bm{k}',\bm{l}} - y^{\alpha,\uparrow\downarrow}_{-\bm{k}',-\bm{k},\bm{q}}\Big) u_{\bm{k}\uparrow} u_{\bm{k}'\downarrow}\\
     &\times v_{\bm{k}\uparrow}v_{\bm{k}'\downarrow}\Big(\langle\alpha^{\dagger}_{\bm{k}\uparrow}\alpha_{\bm{k}\uparrow}\rangle - \langle\alpha^{\dagger}_{\bm{k}'\downarrow}\alpha_{\bm{k}'\downarrow}\rangle\Big) + \Big(x^{\beta,\downarrow\uparrow}_{-\bm{k}',-\bm{k},\bm{q}}\\
     &- x^{\beta,\uparrow\downarrow}_{-\bm{k},-\bm{k}',\bm{l}}\Big)v_{\bm{k}\uparrow}v_{\bm{k}'\downarrow}u_{\bm{k}\uparrow}u_{\bm{k}'\downarrow}\Big( \langle\beta^{\dagger}_{-\bm{k}\uparrow}\beta_{-\bm{k}\uparrow}\rangle - \langle\beta^{\dagger}_{-\bm{k}'\downarrow}\beta_{-\bm{k}'\downarrow}\rangle\Big)\\
     &+\Big(w^{-,\uparrow\downarrow}_{-\bm{k}',-\bm{k},\bm{q}} - z^{+,\uparrow\downarrow}_{-\bm{k},-\bm{k}',\bm{l}}\Big)u_{\bm{k}\uparrow}v_{\bm{k}'\downarrow}v_{\bm{k}\uparrow}u_{\bm{k}'\downarrow}\Big(\langle\alpha^{\dagger}_{\bm{k}\uparrow}\alpha_{\bm{k}\uparrow}\rangle\\
     &+ \langle\beta^{\dagger}_{-\bm{k}'\downarrow}\beta_{-\bm{k}'\downarrow}\rangle + 1\Big) + \Big(w^{-,\downarrow\uparrow}_{-\bm{k},-\bm{k}',\bm{l}}- z^{+,\downarrow\uparrow}_{-\bm{k}',-\bm{k},\bm{q}}\Big)v_{\bm{k}\uparrow}u_{\bm{k}'\downarrow}\\
     &\times u_{\bm{k}\uparrow}v_{\bm{k}'\downarrow}\Big(\langle\beta^{\dagger}_{-\bm{k}\uparrow}\beta_{-\bm{k}\uparrow}\rangle + \langle\alpha^{\dagger}_{\bm{k}'\downarrow}\alpha_{\bm{k}'\downarrow}\rangle + 1\Big)\Bigg\}\\
     &\times c^{\dagger}_{\bm{q}\downarrow}c_{\bm{k} + \bm{q} - \bm{k}'\uparrow}c^{\dagger}_{\bm{l}\uparrow}c_{\bm{k}' + \bm{l}-\bm{k}\downarrow},
\end{aligned}
\end{align}
    
\begin{align}
\begin{aligned}
    &H^{BA} = -\Omega \frac{V^2}{2N}\sum_{\substack{\bm{q}\bm{l}\\\in \Box}}\sum_{\substack{\bm{k}\bm{k}'\\ \in \rm{RBZ}}}\Bigg\{
    \Big(x^{\alpha,\downarrow\uparrow}_{-\bm{k},-\bm{k}',\bm{l}} - x^{\alpha,\uparrow\downarrow}_{-\bm{k}',-\bm{k},\bm{q}} \Big)v_{\bm{k}\downarrow}v_{\bm{k}'\uparrow}\\
    &\times u_{\bm{k}\downarrow}u_{\bm{k}'\uparrow}\Big(\langle\alpha^{\dagger}_{-\bm{k}'\uparrow}\alpha_{-\bm{k}'\uparrow}\rangle - \langle\alpha^{\dagger}_{-\bm{k}\downarrow}\alpha_{-\bm{k}\downarrow}\rangle\Big) + \Big(y^{\beta,\downarrow\uparrow}_{-\bm{k}',-\bm{k},\bm{q}}\\
    &- y^{\beta,\uparrow\downarrow}_{-\bm{k},-\bm{k}',\bm{l}}\Big)u_{\bm{k}\downarrow}u_{\bm{k}'\uparrow}v_{\bm{k}\downarrow}v_{\bm{k}'\uparrow}\Big(\langle\beta^{\dagger}_{\bm{k}'\uparrow}\beta_{\bm{k}'\uparrow}\rangle - \langle\beta^{\dagger}_{\bm{k}\downarrow}\beta_{\bm{k}\downarrow}\rangle\Big)\\
    &+ \Big(z^{-,\downarrow\uparrow}_{-\bm{k},-\bm{k}',\bm{l}} - w^{+,\downarrow\uparrow}_{-\bm{k}',-\bm{k},\bm{q}} \Big)v_{\bm{k}\downarrow}u_{\bm{k}'\uparrow}u_{\bm{k}\downarrow}v_{\bm{k}'\uparrow}\Big(\langle\beta^{\dagger}_{\bm{k}'\uparrow}\beta_{\bm{k}'\uparrow}\rangle\\
    &+ \langle\alpha^{\dagger}_{-\bm{k}\downarrow}\alpha_{-\bm{k}\downarrow}\rangle + 1\Big) + \Big(z^{-,\uparrow\downarrow}_{-\bm{k}',-\bm{k},\bm{q}} - w^{+,\uparrow\downarrow}_{-\bm{k},-\bm{k}',\bm{l}}\Big)u_{\bm{k}\downarrow}v_{\bm{k}'\uparrow}\\
    &\times v_{\bm{k}\downarrow}u_{\bm{k}'\uparrow}\Big(\langle\alpha^{\dagger}_{-\bm{k}'\uparrow}\alpha_{-\bm{k}'\uparrow}\rangle + \langle\beta^{\dagger}_{\bm{k}\downarrow}\beta_{\bm{k}\downarrow}\rangle + 1\Big)\Bigg\}\\
    &\times c^{\dagger}_{\bm{q}\downarrow}c_{\bm{k} + \bm{q} - \bm{k}'\uparrow}c^{\dagger}_{\bm{l}\uparrow}c_{\bm{k}' + \bm{l}-\bm{k}\downarrow}.
\end{aligned}
\end{align}
Here, we have defined, $V \equiv 2\bar{J}/\sqrt{N}$.\\
\indent When considering the ground state expectation value of the boson operator pairs, we only get contributions from $\uparrow$ -bosons with momentum $\bm{k} = 0$, as the ground state is a condensate of $\uparrow$-bosons. Further, restricting to the BCS-case of incoming, as well as outgoing, particles with opposite momenta, the result for the effective interaction is $H_{\rm{pair}} = H^{(1)}_{\rm{pair}} + H^{(2)}_{\rm{pair}}$ where

\begin{align}
\begin{aligned}
    H^{(1)}_{\rm{pair}} &= \frac{V^2}{N}\big(u^2_{0\uparrow} +  v^2_{0\uparrow}\big)n_{0\uparrow}\sum_{\bm{q}\in \Box}\sum_{\substack{\bm{k}\in\\ \rm{RBZ}}}A(\bm{k},\Omega)\\
    &\times\frac{2\omega_{\bm{k}\downarrow}}{(\epsilon_{\bm{q} + \bm{k}} - \epsilon_{\bm{q}})^2 - \omega_{\bm{k}\downarrow}^2}c^{\dagger}_{\bm{q}\downarrow}c_{\bm{q} + \bm{k}\uparrow}c^{\dagger}_{-\bm{q}\uparrow}c_{-\bm{q}-\bm{k}\downarrow},
\end{aligned}
\end{align}

and 

\begin{align}
\begin{aligned}
    H^{(2)}_{\rm{pair}} &= \frac{V^2}{N}\sum_{\substack{\bm{q} \\\in \Box}}\sum_{\substack{\bm{k}\bm{k}'\\ \in \rm{RBZ}}}B(\bm{k},\bm{k}',\Omega)\\
    &\times\frac{2\big(\omega_{\bm{k}\uparrow} + \omega_{\bm{k}'\downarrow}\big)}{\big(\epsilon_{\bm{q} + \bm{k} - \bm{k}'} - \epsilon_{\bm{q}}\big)^2 - \big(\omega_{\bm{k}\uparrow} + \omega_{\bm{k}'\downarrow} \big)^2}\\
    &\times c^{\dagger}_{\bm{q}\downarrow}c_{\bm{q} + \bm{k} - \bm{k}'\uparrow}c^{\dagger}_{-\bm{q}\uparrow}c_{- \bm{q} + \bm{k}' - \bm{k}\downarrow}.
\end{aligned}
\end{align}
The functions $A$ and $B$ are defined in the main text. The contributions to $H^{(1)}_{\rm{pair}}$ come from the expectation value of the $\uparrow$-bosons with momentum $\bm{k} = 0$, $n_{0\uparrow}$, while the contributions to $H^{(2)}_{\rm{pair}}$ originate with the terms without boson operators. Moving the contributions from $H^{(2)}_{\rm{pair}}$ where $\bm{k} = 0$ over to $H^{(1)}_{\rm{pair}}$, and using that $u^2_{0\uparrow}, v^2_{0\uparrow} \gg 1$, we can rewrite $H^{(1)}_{\rm{pair}}$ as 

\begin{align}
\begin{aligned}
    &H^{(1)}_{\rm{pair}} = V^2\Tilde{Q}_0\sum_{\bm{q}\in \Box}\sum_{\substack{\bm{k}\in\\ \rm{RBZ}}}A(\bm{k},\Omega)\\
    &\times\frac{2\omega_{\bm{k}\downarrow}}{(\epsilon_{\bm{q} + \bm{k}} - \epsilon_{\bm{q}})^2 - \omega_{\bm{k}\downarrow}^2}c^{\dagger}_{\bm{q}\downarrow}c_{\bm{q} + \bm{k}\uparrow}c^{\dagger}_{-\bm{q}\uparrow}c_{-\bm{q}-\bm{k}\downarrow},
\end{aligned}
\end{align}
where $\Tilde{Q}_0$ is a quantity of order unity, closely related to the sublattice magnetization. 

\bibliographystyle{apsrev4-1}  
\addcontentsline{toc}{chapter}{\bibname}
\bibliography{main}  

\end{document}